\documentclass{pnastwo}

\usepackage{graphicx}
\usepackage{epstopdf}

\usepackage{amssymb,amsfonts,amsmath}
\usepackage{rotating}
\usepackage{xr}
\usepackage{color}
\usepackage{url}
\usepackage{rotating}

\begin{document}

\title{Interspecific Introgressive Origin of Genomic Diversity in the House Mouse} 

\author{Kevin J. Liu\affil{1}{Department of Computer Science and Engineering, Michigan State University, East Lansing, MI 48824 (Research was performed while KJL was at Rice University)},
 Ethan Steinberg\affil{2}{Department of Computer Science, Rice University, Houston, TX 77005},
 Alexander Yozzo\affil{2}{},
 Ying Song\affil{3}{Biosciences, Rice University, Houston, TX 77005},
 Michael H. Kohn\affil{3}{}, 
 \and Luay Nakhleh\affil{2}{}\affil{3}{}}

\contributor{Submitted to Proceedings of the National Academy of Sciences
of the United States of America}

\maketitle

\begin{article}

\begin{abstract}
We report on a genome-wide scan for introgression in the house mouse ({\em Mus musculus domesticus}) involving the 
Algerian mouse
 ({\em Mus spretus}), using samples from the ranges of sympatry and allopatry in Africa and Europe.  Our analysis reveals wide variability 
 in introgression signatures along the genomes, as well as across the samples. We find that fewer than half of the autosomes 
 in each genome harbor all detectable introgression, while the X chromosome has none. Further, European mice carry more {\em M. spretus} alleles than the sympatric African ones. Using the length distribution and sharing patterns of 
 introgressed genomic tracts across the samples, we infer, first, that at least three distinct 
 hybridization events involving {\em M. spretus} have occurred, one of which is ancient, and the other two are recent (one presumably due to warfarin rodenticide selection). Second, several of the 
 inferred introgressed tracts contain genes that are likely to confer adaptive advantage. Third, introgressed tracts might contain 
 driver genes that determine the evolutionary fate of those tracts. Further, functional analysis revealed introgressed genes that 
 are essential to fitness, including the {\em Vkorc1} gene, which is implicated in rodenticide resistance, and olfactory receptor genes.  
 Our findings highlight the extent and role of introgression in nature, and call for careful analysis and interpretation of house 
 mouse  data in evolutionary  and genetic studies. 
\end{abstract} 

\keywords{{\em Mus musculs} | {\em Mus spretus} | hybridization | adaptive introgression | PhyloNet-HMM}

\section{Significance Statement}
The mouse has been one of the main mammalian model organisms used for genetic and biomedical research. 
Understanding the evolution of house mouse genomes would  shed light not only on genetic interactions and their 
interplay with traits in the mouse, but would also have significant implications for human genetics and health. Analysis using  
a recently developed statistical method shows that the house mouse genome is a mosaic that contains previously 
unrecognized contributions from a different mouse species. We traced these contributions to ancient and recent interbreeding  
 events. Our findings reveal the extent of introgression in an important mammalian genome, and provide an 
approach for genome-wide scans of introgression in other eukaryotic genomes.  
\vspace{0.2in}

\begin{figure}[ht]
\begin{center}
\centerline{\includegraphics[scale=0.3]{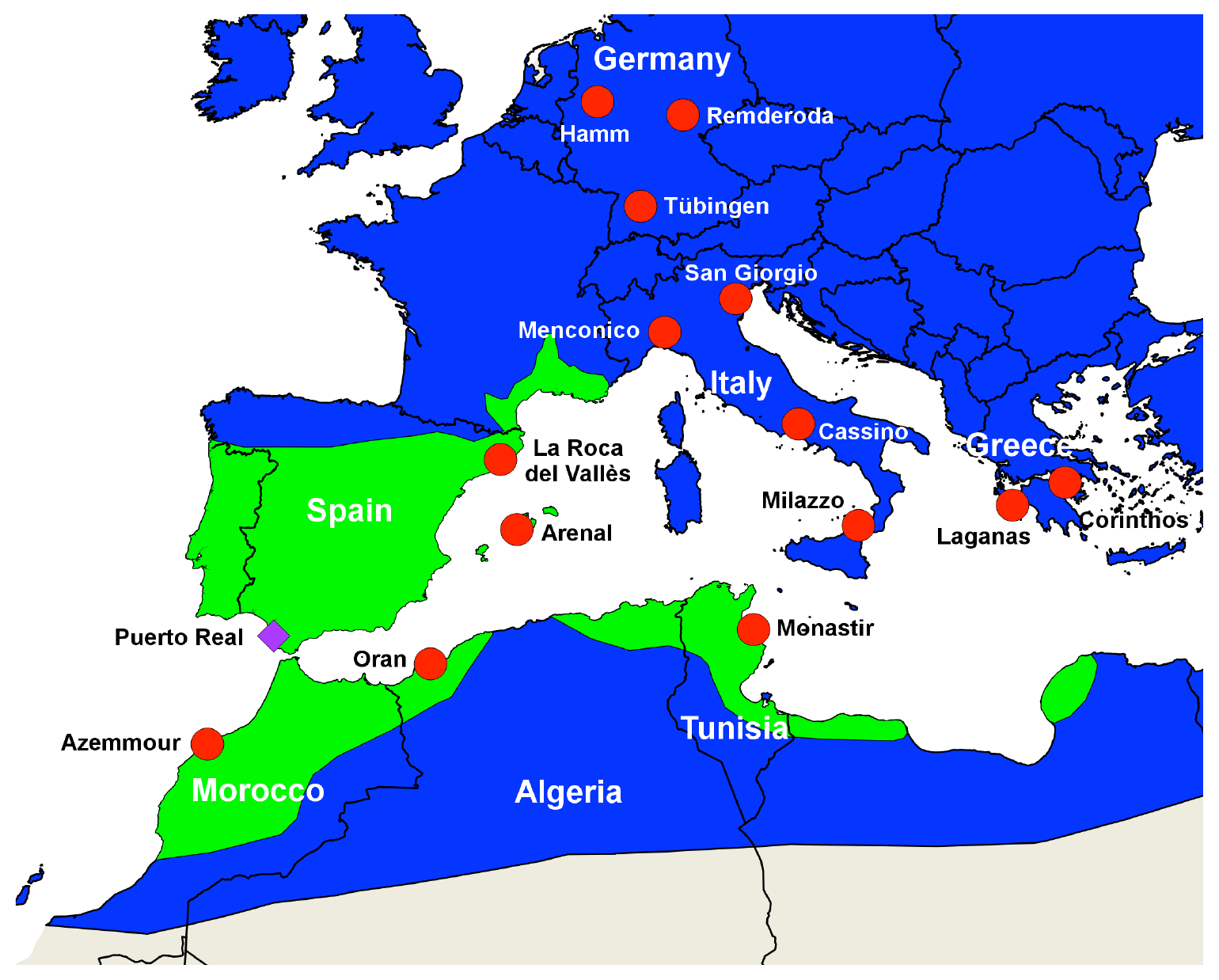}}
\end{center}
\caption{
{\bf Population ranges and samples used in our study.} 
The population range of {\it M. spretus} is shown
in green \cite{mspretus}, and the population range
of {\it M. m. domesticus} includes the blue regions,
the range of {\it M. spretus}, and beyond \cite{Guenet200324}.
Wild {\it M. m. domesticus} and {\it M. spretus}
samples were obtained from locations marked with red circles and 
purple diamonds, respectively.
The samples originated from 
within and outside the area of sympatry between the two species.
(See Supplementary Table \ref{samples} 
for additional details about the samples used in our study.)
\label{sample-gis}}
\end{figure}

\dropcap{C}lassical laboratory mouse strains, as well as newly established wild ones are widely used by 
geneticists for answering a diverse array of questions \cite{Guenet200324}.  Understanding the 
genome contents and architecture of these strains is very important for studies of natural variation 
and complex traits, as well as evolutionary studies in general \cite{Yang2007}. {\em Mus spretus}, a sister species of {\em M. musculus}, impacts the findings in {\em M. musculus} investigations for at least two reasons. First, it was 
deliberately interbred with  laboratory {\em M. musculus} strains to introduce genetic variation \cite{spretuscross}. Second, {\em M. m. domesticus}
 is partially sympatric (naturally co-occurring) with {\em M. spretus} (Fig. \ref{sample-gis}). 

Recent studies have examined
admixture between subspecies of house mice \cite{Yang2011,Keane2011,Staubach2012,Teeter01012008},
but have not studied introgression with {\it M. spretus}.
In at least one case \cite{Yang2011},
the introgressive descent origins of the mouse genome 
were hidden due to data post-processing
that masked introgressed genomic regions as missing data. In another study reporting whole-genome sequencing of 17 classical
lab strains \cite{Keane2011}, {\it M. spretus} was used 
as an outgroup for phylogenetic analysis. The authors were surprised to find that 12.1\%
of loci failed to place
{\it M. spretus} as an outgroup to the {\it M. musculus} clade.
The authors concluded that {\it M. spretus} was not a reliable outgroup, but did
not pursue their observation further.
 On the other hand,  in a 2002 study \cite{FrenchStudy}, Orth {\em et al.} compiled data on allozyme, microsatellite, and mitochondrial  variation in house mice from Spain (sympatry) and nearby countries in Western and Central Europe. Interestingly then, allele sharing between the species was observed in the range of sympatry but not so outside in the range of allopatry. The studies demonstrated the possibility of natural  hybridization between these two sister species. Further, the study of Song {\em et al.}  \cite{Song20111296} demonstrated a recent adaptive introgression from {\it M. spretus} 
into some {\it M. m. domesticus} populations in the wild, involving the vitamin K epoxide reductase subcomponent 1 ({\it Vkorc1}) gene, which was later shown to be more widespread in Europe albeit geographically restricted to parts of southwestern and central Europe \cite{PS:PS2254}. 

 Major, unanswered questions arise from these studies. First, is the vicinity around the {\it Vkorc1} gene 
an isolated case of adaptive introgression in the house mouse genome, or do many other such regions exist?
 Second, is introgression from {\em M. spretus} into {\em M. m. domesticus} common outside the range of sympatry? Third, have there been other hybridization events, and in particular, more ancient ones? Fourth, what  role do introgressed
genes, and more generally, genomic regions, play? 

To investigate these open questions, we used genome-wide variation data from 20 {\em M. m. domesticus} samples (wild and wild-derived) from the ranges of sympatry and allopatry, as well as  two {\em M. spretus} samples. 
 For detecting introgression, we used PhyloNet-HMM \cite{Liu2014}, a newly developed method for statistical inference of introgression in genomes 
while accounting for other evolutionary processes, most notably incomplete lineage sorting (ILS). 

 Our analysis provides answers to the  questions posed above. First, we find signatures of introgression in each of the {\em M. m. domesticus} samples. The amount of introgression varies across the autosomes of each genome, with a few chromosome harboring all detectable introgression, and most of the genomes have none. We detected no introgression on the X chromosome.  Further, the 
 amount of introgression varied widely across the samples. Our analyses demonstrate introgression outside the range of sympatry. 
 In fact, our results show more signatures of introgression in the genomes of samples from Europe than in those of samples from 
 Africa. For the third question, we utilized the length distribution and sharing patterns of introgressed regions across the samples 
 to show support for at least three hybridization events: an ancient hybridization event that predates the colonization of Europe by {\em M. m. domesticus} and two more recent events, one of which presumably occurred about 50 years ago and is related  to warfarin resistance selection \cite{Song20111296}. For the fourth question, our functional analysis of the introgressed genes 
 shows enrichment for certain categories, most notably olfaction---an essential trait for the fitness of rodents. Understanding the 
 genomic architecture and evolutionary history of the house mouse has broad implications on various aspects of evolutionary, 
 genetic, and biomedical research endeavors that use this model organism. The PhyloNet-HMM method \cite{Liu2014} can 
 be used to detect introgression in other eukaryotic species, further broadening the impact of this work.

\section{Results}
We now describe our findings of introgression within the individual genomes, as well as across the genomes of the 20 {\em M. m. domesticus} samples (40 haploid genomes). The four African samples as well as the two samples from Spain are sympatric with {\em M. spretus}, whereas the other samples are allopatric (Fig. \ref{sample-gis}). 

\begin{figure}[!ht]
\begin{center}
\centerline{\includegraphics[scale=0.6]{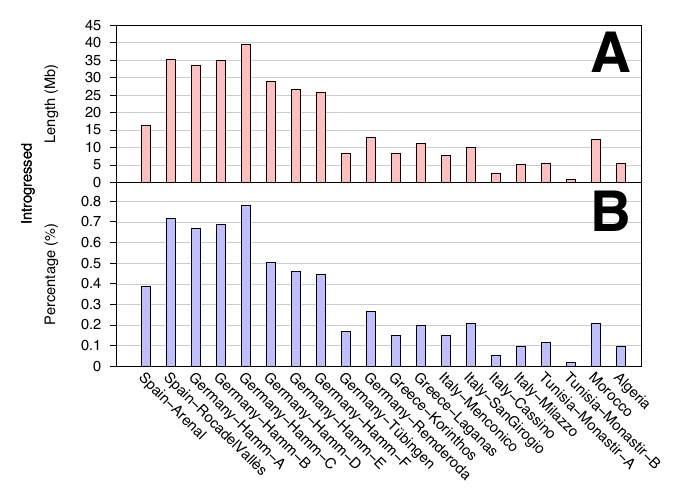}}
\end{center}
\caption{{\bf Amount of introgressed genetic material in the 20 {\em M. m. domesticus} samples.} 
 (A) The amount of introgressed genetic material in Mb per sample. (B) The amount of introgressed genetic material as a percentage of the genome length per sample. 
\label{fig:dist}}
\end{figure}

\paragraph{Genome-wide signals of introgression.}
Our analysis detected introgression in the genomes of all 20 {\em M. m. domesticus} samples; Fig.~\ref{fig:dist} (see SI Appendix for complete scans of introgression of all 20 samples). 
 However, the patterns of introgression varied  across the chromosomes within each individual genome, as well as across the genomes. In terms of within-genome variability, a few chromosomes in each genome carried almost all of the introgressed regions. For example, all detected introgressed regions resided on five chromosomes in the sample from Spain-RocadelVall\`{e}s.  For all samples, fewer than half the chromosomes of a sample's genome carried any detected introgression (Figures S\ref{hybrid-chr1}--S\ref{hybrid-chr19} in the SI Appendix). The analysis did not detect any introgression on chromosome X  (Fig. S\ref{hybrid-chrX} in the SI Appendix). Further, in the two samples from Spain and the six 
Germany-Hamm samples, one or two chromosomes carried over 50\% of all detected introgression. Generally, the percentage of introgressed sites in a genome ranged from about 0.02\% in a sample from Tunisia to about 0.8\% in samples from Germany (Fig.~\ref{fig:dist}). The large extent of detected introgression between {\em M. spretus} and {\em M. m. domesticus} seen on chromosome 17 in the samples from Spain (see SI Appendix) 
 merits further investigation. The introgressed regions  spatially coincide with the known polymorphic 
 recombination-suppressing inversions and t-hapolotypes in house mice  \cite{hammer1989evolution}.
 
 The amount of introgression in the genomes of the 20 samples points qualitatively to three groups of samples: Group I, which includes the two samples from Spain and the six Germany-Hamm samples; Group II, which includes the two other Germany samples and the Italy and Greece samples; and, Group III, which includes the samples from Africa. Variability in the amount of introgresison across samples within each group is much smaller than that across groups, as is the amount of sharing of introgressed regions. Further, Group I has the most introgression and Group III has the least. Notice that all samples within Group I, except for the one from Spain-Arenal, contain the
 introgression from  
 {\em M. spretus} that carries {\em Vkorc1} \cite{Song20111296}. Group II contains all the allopatric European mice that do not carry {\em Vkorc1}, and Group III contains all the sympatric African samples.  This categorization guides our results displays and analyses below. These results answer the first two questions we posed above in the affirmative: there are introgressed genomic regions beyond the region that contains {\em Vkorc1}, and introgression is present in all 20 samples, pointing to potential hybridization outside the range of sympatry. Quantifying how common such hybridization is outside the range of sympatry, though, requires denser sampling that is beyond the scope of this work. 

\paragraph{Support for more than a single hybridization event.} 
To answer the third question of whether multiple, distinct hybridization events involving {\em M. spretus} and {\em M. m. domesticus} have occurred, we focused on two analyses: Inspecting the introgressed tract length distribution, where an introgressed tract is defined as a maximally contiguous introgressed region, and inspecting the sharing pattern of introgressed regions across the samples. Repeated back-crossing, recombination, and drift result in fragmentation of introgressed regions, with very long regions pointing to recent hybridization events. On the other hand, selection on an adaptive introgressed regions could also maintain it for long periods, confounding the tract length-based analysis of the age of hybridization. However, if a long region is shared across some, but not all, samples from the population, that increases the likelihood of a recent hybridization hypothesis. 

\begin{figure*}[!ht]
\begin{center}
\centerline{\includegraphics[scale=0.45]{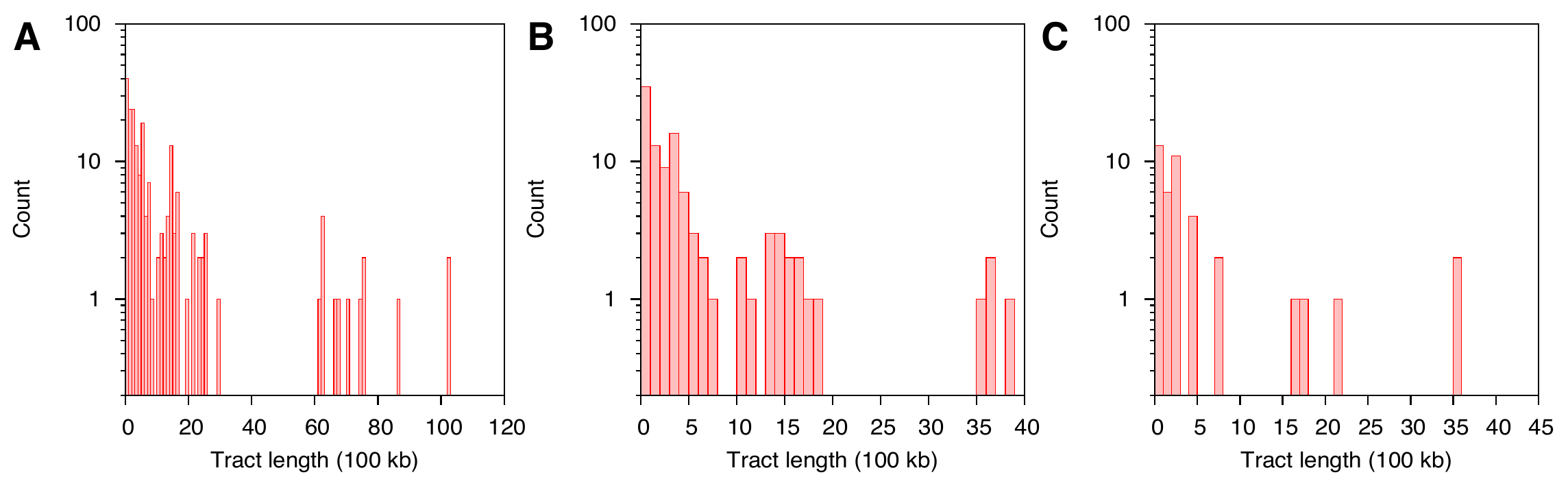}}
\end{center}
\caption{{\bf Distributions of introgressed tract lengths detected in the 40 haploid genomes.} (A) Group I: The six Germany-Hamm samples and two Spain samples. (B) Group II: The four Italy, two Greece, and two other Germany samples. (C) Group III: The Algeria, Morocco, and two Tunisia samples. Note the x-axis scale difference between panel A and the other two panels. (See main text for the rationale behind the grouping.) 
\label{fig:tractlength}}
\end{figure*}

\begin{figure}[!ht]
\begin{center}
\centerline{\includegraphics[scale=0.33]{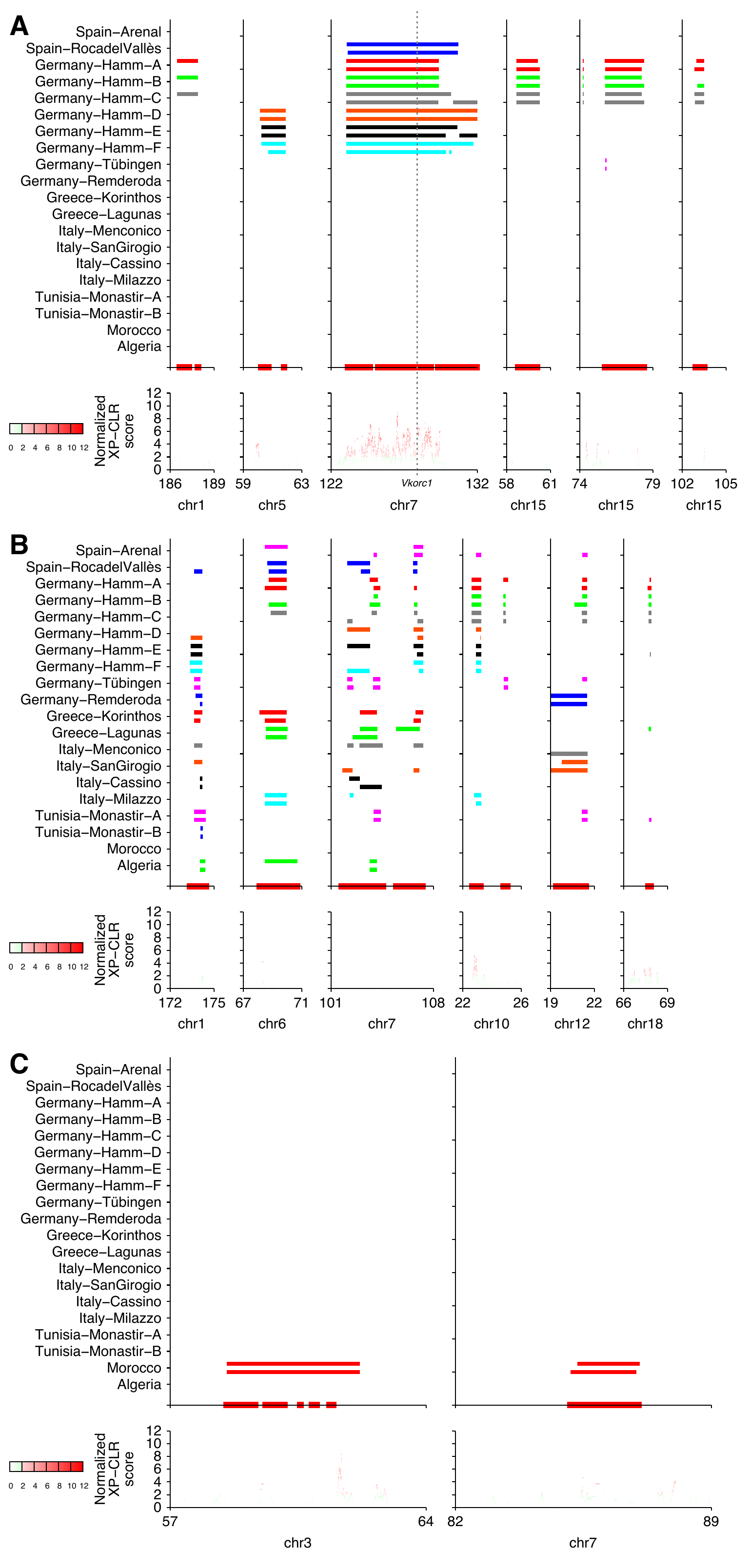}}
\end{center}
\caption{{\bf Three different introgression patterns across the 20 {\em M. m. domesticus} samples.} 
(A) Introgressed regions that are exclusive to the Germany-Hamm and Spain samples. (B) Introgressed regions that 
are shared across the samples. (C) Introgressed regions that are exclusive to African samples. For each sample, scans 
from both haploid chromosomes are shown. A posterior decoding cutoff of 95\% was used to declare a site introgressed (see 
main text for more details). The red squares on the x-axis of the top part of each panel denotes the locations of genes in 
introgressed regions (given the scale, the squares appear overlapping, but the genes are not overlapping). The location of {\em Vkorc1} on chromosome 7 is indicated with a dashed vertical line in (A). The bottom 
part of each panel shows selective sweep statistics, which are 
normalized XP-CLR scores \cite{Chen01032010} based on a comparison of rodenticide-resistant to 
rodenticide-susceptible wild {\em M. m. domesticus} samples. \label{fig:introgressedcharts}}
\end{figure}

For each of the three groups, we plotted the distribution of introgressed tract lengths, where an introgressed tract is defined as a maximally contiguous introgressed region (Fig.~\ref{fig:tractlength}).  
 The figure shows that Group I contains the only samples that have introgressed tracts of lengths exceeding 4 Mb. All these tracts correspond to the adaptively introgressed region that contains {\em Vkorc1} between positions 122 Mb and 132 Mb on chromosome 7 (Fig.~\ref{fig:introgressedcharts}A). The exclusivity of these very long introgressed tracts to Group I points to a very recent 
hybridization event involving {\em M. spretus}, in agreement with the assessment of \cite{Song20111296}. Except for this group of 
introgressed tracts, the three distributions are very similar, with an excess of very short tracts and a smaller number of longer tracts 
(up to 4 Mb). The very short tracts could be a signal of ancient hybridization, or just incorrect inference by the method (detecting very short introgressed regions is very hard due to low signal-to-noise ratio).  However, the pattern of sharing of introgressed regions across the samples supports a hypothesis of ancient hybridization, as we now discuss. 

Fig.~\ref{fig:introgressedcharts} shows examples of three different patterns of introgression across the samples (full genome-wide scans of all samples can be found in the SI Appendix). 
 Fig.~\ref{fig:introgressedcharts}A shows introgressed tracts that are shared exclusively among samples in Group I (we hypothesize 
 that the Spain-Arenal sample underwent a secondary loss of the introgressed {\em Vkorc1}-containing region that it once had). As we discussed 
above, these point to at least one, very recent hybridization event involving {\em M. spretus}. Fig.~\ref{fig:introgressedcharts}B shows introgressed tracts that are shared across samples from all three groups. This pattern points to an ancient hybridization event involving {\em M. spretus} and that precedes the ancestor of all {\em M. m. domesticus} samples in the study. It is important to note here that this pattern could also be a signature of balancing selection on standing variation prior to the split of {\em M. musculs} and {\em M. spretus}. We discuss this possibility in the Discussion section below. Fig.~\ref{fig:introgressedcharts}C shows introgressed regions, of considerable length, in the sample from Morocco. This is, again, a signature of a recent hybridization event that is different from that involving the large tracts on chromosome 7 in Group I. 

Putting together all the evidence, the data supports a hypothesis of at least three distinct hybridization events. One hybridization event   is ancient, predating the colonization of Europe by {\em M. m. domesticus} upwards of 2,000 years ago \cite{auffray1992did}.  The other two hybridization events are more recent, and one of them presumably occurred about 50 years ago and is related  to warfarin resistance selection \cite{Song20111296}. 

\paragraph{Adaptive signals of introgression.}
Introgressed genomic tracts and the genes they carry are generally assumed to be neutral or deleterious. Further, such tracts would naturally be expected to 
be present in the genomes of sympatric hybridizing taxa. Consequently, in the samples we considered here, one would expect to 
find more introgressed tracts, if any, in the sympatric mice (the African and Spanish samples) than in the allopatric ones. However, our 
results give a very different picture from these expectations. We hypothesize that some of these introgressed tracts have conferred selective advantage on the mice that carry them. For example, the introgressed region on chromosome 
7 in Group I contains the {\em Vkorc1} gene whose introgression and adaptive role were discussed in the context of warfarin resistance selection in 
\cite{Song20111296}. To identify whether other introgressed regions of adaptive role are associated with that {\em Vkorc1}-containing region, 
we applied the selective sweep measure of \cite{Chen01032010} to a comparison of rodenticide-resistant to
rodenticide-susceptible wild {\it M. m. domesticus} samples, which favors detection of the recent rodenticide-related
selective sweep (within the last $\sim 50$ years).
Not surprisingly, the selective sweep statistics 
in the {\it Vkorc1}-containing region were among the largest of any 
detected in our study; Fig \ref{fig:introgressedcharts}A (see SI Appendix for full results).
We also detected selective sweeps outside the {\it Vkorc1} region.

\begin{figure}[!ht]
\begin{center}
\centerline{\includegraphics[scale=0.35]{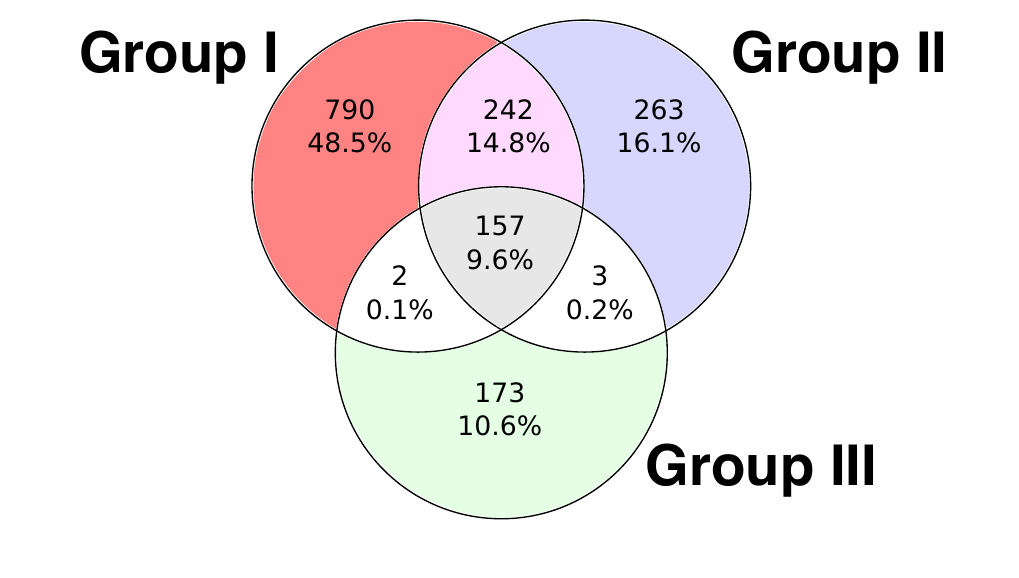}}
\end{center}
\caption{{\bf Venn diagram of the three sets of genes in introgressed regions in Groups I, II, and III of samples.}  Introgression was called  based on
a posterior decoding probability cutoff
of 95\%. For each circle in the Venn diagram,
two quantities are shown: (top) the number of genes, and (bottom) the percentage
of all introgressed genes found in our study.
\label{fig:venndiagram}}
\end{figure}

To assess the potential adaptive benefit of other introgressed regions, we utilized the frequencies of the introgressed regions, as 
reflected by the sharing patterns.  For example, the shared introgressed regions across sympatric and allopatric samples on chromosome 1 and 7, as shown in Fig. \ref{fig:introgressedcharts}B, point to a hypothesis of adaptive roles of parts of these 
regions. To further zoom in on these shared regions, we analyzed the sharing patterns of genes across in the introgressed regions 
across all samples. Fig. \ref{fig:venndiagram} shows the Venn diagram of the sets of introgressed genes in Groups I, II, and III. 

Fig. \ref{fig:venndiagram} shows that the two European groups have 399 introgressed genes in common, almost twice the number of introgressed 
genes that are in common between either of them and the African group. We hypothesize that the set of 157 genes that are shared 
across all three groups contain a subset that we call ``driver genes"---those that have driven the maintenance of those introgressed 
regions for a long time across the samples. In our proposed classification, driver genes would be  beneficial upon introgression and would be subject to selection. Genes that are introgressed in one group, but not the others, are potential 
 neutral, linked ``passenger" genes. While passenger genes would be expected to be neutral, they could also introduce new 
 polymorphisms into {\em M. m. domesticus} genomes and could become subject to selection at some point during its sojourn time. 

Individual driver genes on introgressed tracts are not expected to result in functional enrichment scores. For example, the introgressed region that contains {\em Vkorc1} on chromosome 7 has many other genes, yet is not enriched for any functional categories. On 
the other hand, introgressed tracts with an abundance of genes from a given family tend to result in significant enrichment scores of the tracts. We illustrate this with two examples related to olfaction, a multi-genic trait known to be essential for the fitness of rodents (full lists of the genes in introgressed regions and their Gene Ontology functional enrichment are given in the SI Appendix).  
The introgressed tract on chromosome 1 that is shared by mice from Africa and Europe, including allopatric mice, is significantly enriched (P= 5.9E-8 after Benjamini-Hochberg \cite{BenjaminiHochberg} correction) for genes involved in olfactory transduction and encode olfactory receptor genes (13 out of 36 genes) located on the contiguous tracts (Fig. \ref{fig:introgressedcharts}B). It is conceivable that this group of genes, or a subset of these, may have acted as a driver for this introgressed fragment carrying at least another set of 23 passenger genes.
 Similarly, the region on chromosome 7 shared by the African samples is highly enriched (P=3.6E-6) for genes involved in olfactory transduction  and encode at least 15 olfactory receptors amongst at least 62 genes situated on this tract. Evidently, large tracts have become polymorphic for introgressed and native repertoires of olfactory receptor alleles.

\section{Discussion}
The biological significance of hybridization and introgression in the evolution of new traits in natural eukaryotic populations has ignited much research into these two processes \cite{arnold2004transfer}. Introgressed genetic material can be neutral and go unnoticed in terms of phenotypes but can also be adaptive and affect phenotypes
 \cite{Song20111296,schmidt1977differences}. Notably, these processes have played a crucial role in the domestication of plants and animals, and appear to be common in natural populations of plants \cite{arnold2004transfer}.
 For example, the importance of introgression has become a central discussion point when reconstructing the evolution of primates, including humans \cite{arnold2006natural}. Further, it has now become clear that the genomes of model organisms prior to their adoption as laboratory models by humans have been shaped by hybridization and introgression in their natural ancestral populations, such as in mice and macaques  \cite{FrenchStudy,stevison2008determining,osada2010ancient}. Such influx of genetic variation of inter-subspecific or inter-specific origins is expected to continue, as wild-derived strains of mice will contribute to the collaborative cross in laboratory mice, and Primate Research centers continue to rely on imports of macaques from Asia. 

Classical laboratory mouse strains, as well as newly established wild ones are widely used by 
geneticists for answering a diverse array of questions \cite{Guenet200324}.  Understanding the 
genome contents and architecture of these strains is very important for studies of natural variation 
and of complex traits, as well as evolutionary studies in general \cite{Yang2007}.
 For example, large-scale efforts have been made to decode the genetic background of most commonly used laboratory mouse strains, including inbred and wild-derived strains of {\em M. m. domesticus}, and of other other subspecies of the laboratory mouse, including {\em M. m. musculus} and {\em M. m. castaneus} \cite{Yang2007}. Amongst the numerous insights of the evolutionary genomic analyses of the laboratory mouse and its wild relatives were that inter-subspecific introgression between strains has been common \cite{Yang2007}. 
  In addition to understanding the ancestry and mosaic structure of laboratory mouse genomes, detecting introgression is also of biomedical significance. In a recent study \cite{Song20111296}, the authors discovered in a mouse model resistance  to the commonly used anticoagulant warfarin \cite{scully2002warfarin} through the acquisition of a mutated version of a 
 key enzyme of the vitamin K cycle, {\em Vkorc1}, that is targeted by warfarin. 
  While previous genome-wide studies in mice 
focused on polymorphism and introgression within the {\em M. musculus} group \cite{Yang2007,Yang2011}, we focused here 
on introgression of {\em M. spretus} genomic material into the genomes of several {\em M. m. domesticus} samples from
the regions of sympatry and allopatry in Africa and Europe.  These analyses are now enabled by our recently developed method 
for statistical inference of introgression in the presence of other evolutionary events, most notably incomplete lineage sorting \cite{Liu2014}. 

In terms of the debate surrounding the importance of introgression in animal evolution an important result of our genome-wide study is that it is not only a recent strong and potentially human-driven selection (warfarin rodenticide
and {\em Vkorc1}) that has promoted introgression in natural populations of mice. Hybridization and introgression between {\em M. spretus} and {\em M. m. domesticus} appear to be  natural processes 
spanning at least several thousand years. Nevertheless, even from a rather dense genome-wide survey such as ours it is difficult to discern how frequent introgression occurs. This is because most hybridization does not lead to introgression, as drift and selection tend to remove introgressed regions. However, here we infer at least three hybridization events, one in the distant past and two of more recent timing (including the introgression of {\em Vkorc1}).  We suspect that this is an underestimate of the frequency of hybridization and introgression between {\em M. spretus} and {\em M. m. domesticus} in the wild since the species have established secondary contact a few thousand years ago when house mice reached the Mediterranean Basin on their westward spread into Northern Africa and Europe.

In terms of a role of selection on introgressed tracts, our genome-wide scan revealed informative patterns. First, 
introgression is limited to a few autosomes and absent from the X chromosome. This is consistent with a strong role of purifying selection and drift in the removal of introgressed material. We find it noteworthy that tracts are very frequently found in the homozygous state, which indicates that introgressed variants can be recessive as well as dominant. The sharing of tracts across samples is consistent with positive selection on introgressed material (adaptive introgression). Such tract sharing is observed over long time scales, such as for chromosome 1, as well as over shorter time scales and locally, as is suggested by tract sharing by subsets of samples from presumably local populations such as Hamm in Germany or African samples. Finally, it is common to infer that introgressed regions are adaptive if these are found outside the area of sympatry. As we observed numerous introgressed tracts, both presumably old and young as judged by their tract lengths, it is reasonable to assume that selection might have favored the spread of some of these variants into the allopatric range of house mice. 

Selection could have confounding effects on our analyses and inferences. Specifically, it is important to note that various evolutionary events could give rise to genomic patterns and signals that resemble those 
created by introgression, including incomplete lineage sorting (ILS), convergence, and ancestral polymorphism coupled with 
 balancing selection \cite{hedrick2013adaptive}. Indeed, all these processes could confound the detection of introgression 
 \cite{nakhleh2013computational}. The introgression detection method \cite{Liu2014} that we use here accounts for ILS (hence, 
 ancestral polymorphism that is not under balancing selection), and 
 employs finite-site models in a statistical inference framework, which helps account for convergence at the nucleotide level. Currently, methods that automatically distinguish introgression from balancing selection do not exist however. For example,  
 recent studies of adaptive introgression \cite{Green07052010,Heliconius2012} focused solely on introgression. 
 Our  current analysis of multiple {\em M. m. domesticus} from the ranges of sympatry and allopatry in Africa and Europe mostly point to a very polymorphic signal 
 of introgression across these samples which, consequently, decreases the likelihood that ancestral polymorphism with balancing 
 selection acting on it could explain the patterns we see in the data.  
 We also scanned genomes of samples from  {\em M. m. musculus}, which is a sister subspecies of {\em M. m. domesticus} (Fig.  
  S\ref{mmm-control-chr7} in the SI Appendix). As the figure shows, no introgression was detected in the {\em M. m. musculus} sample (a similar result to that shown in \cite{Liu2014}), which further weakens the plausibility of balancing selection acting from the most recent common ancestor of {\em M. musculus} and {\em M. spretus} until the present day.
 Furthermore, many of the introgressed regions we detect are much 
  longer than regions with balancing selection signatures reported in previous studies. For example, in humans, DeGiorgio {\em et al.}   reported that the maximum contiguous genomic region with balancing selection signal was of length $\sim$40 kb (near the {\em FANK1} gene), and which ``surprised" them because it was ``abnormally large for balancing selection" \cite{degiorgio}. Introgressed 
  tract lengths on the order of megabases would require much stronger balancing selection than have been previously reported in the literature.  Still, for some of the shared introgressed regions, we cannot rule out the possibility of balancing selection, as an alternative to introgression. While our method for detecting introgression can be confounded by balancing selection, a similar issue might 
   arise for methods that detect balancing selection. For example, DeGiorgio {\em et al.} noted recently that gene flow is 
   very likely to confound their test for balancing selection \cite{degiorgio}. New methods need to be developed to account for the two processes simultaneously, as detecting balancing selection could shed light on the evolution of species \cite{leffler2013multiple}. 

Adaptive introgressive hybridization may be an important source for novel functional genetic variants, and combinations thereof, that encode novel traits upon which selection could act. Here, one objective of the study is to discern whether multiple introgressed genes encode the previously reported warfarin resistance trait. Our functional annotation of the introgressed tract did not reveal any discernible enrichment for genes that clearly modulate warfarin resistance. Moreover, we did not observe a patter where all mice that carry the {\em Vkorc1} introgression share introgressed tracks of comparable length. We therefor conclude, that, until further samples are analyzed and tracts are annotated in terms of function more comprehensively, the adaptive introgressive hybridization leading to warfarin resistance in house mice requires the introgression of only the {\em Vkorc1}. Whether the introgressed {\em Vkorc1} interacts with the native genes in pathways resulting in warfarin resistance cannot be deduced from our analysis at this time.

We observed that the raw material for a complex trait known to be important to rodent life history could be introgressed. We noticed the large numbers of olfactory receptor genes for which now polymorphic and divergent copies segregate in the populations of house mice \cite{Young15052002}. 
It is known that both minor nucleotide differences as well as larger scale differences in the olfactory receptor repertoire have measurable phenotypic consequences in mice  \cite{Young15052002}. 
Thus, we hypothesize that wild mice from Europe that have experienced gene flow, such as seen for chromosome 1 enriched for olfactory receptor clusters, indeed may be the subject of natural selection.

\begin{materials}
Our study utilized two {\em M. spretus} samples and twenty {\em M. m. domesticus} samples from the ranges of sympatry and allopatry, that were either newly sampled or from previous publications. For details on compiling the sequence data, genotyping and phasing it, as well as single nucleotide polymorphism (SNP) calling, see the SI Appendix. 
 We used PhyloNet-HMM \cite{Liu2014} to scan
{\em M. musculus} genomes for 
segments with introgressed origin from {\em M. spretus}.
For every haploid genome {\em M. m. domesticus}, we analyzed it along with the the {\em M. spretus} genomes and 
detected introgression. See the SI Appendix for full details on how the analysis was done. 
 We used XP-CLR \cite{Chen01032010} version 1.0 to scan for selective sweep patterns.
The method was run using its default settings. See the SI Appendix for the set of samples used in
these scans.

\end{materials}

\begin{acknowledgments}
We thank Yun Yu for help with the PhyloNet-HMM software. 
The work was partially supported by R01-HL091007-01A1 to MHK from the National Institutes of Health, 
National Heart, Lung and Blood Institute, and by startup funds from Rice University to MHK.
LN was supported in part by NSF grants DBI-1062463 and CCF-1302179, and grant R01LM009494 from the National Library of Medicine (NLM). 
KJL was partially supported by a training fellowship from the Keck Center of the Gulf Coast
Consortia, on the NLM Training Program in Biomedical Informatics,
NLM T15LM007093. 
 \end{acknowledgments}

%

\end{article}


\end{document}


\title{Supplementary Material: Interspecific Introgressive Origin of Genomic Diversity in the House Mouse} 

\author{Kevin J. Liu\affil{1}{Department of Computer Science and Engineering, Michigan State University, East Lansing, MI 48824 (Research was performed while KJL was at Rice University)},
 Ethan Steinberg\affil{2}{Department of Computer Science, Rice University, Houston, TX 77005},
 Alexander Yozzo\affil{2}{},
 Ying Song\affil{3}{Biosciences, Rice University, Houston, TX 77005},
 Michael H. Kohn\affil{3}{}, 
 \and Luay Nakhleh\affil{2}{}\affil{3}{}}

\contributor{Submitted to Proceedings of the National Academy of Sciences
of the United States of America}

\maketitle

\begin{article}

\section{Supplementary Materials and Methods}

Our study utilized 29 mice that were either newly sampled or from previous
publications (Supplementary Table \ref{samples}).
The newly sampled mouse was obtained as part of a tissue sharing agreement 
between Rice University and Stefan Endepols at Environmental Science, Bayer CropScience AG, D-40789 Monheim, Germany 
and Dania Richter and Franz-Rainer Matuschka at Division of Pathology, Department of Parasitology, 
Charit\'{e}-Universit\"{a}tsmedizin, D-10117 Berlin, Germany
(reviewed and exempted by Rice University IACUC).
 We included wild {\em M. m. domesticus} samples from the sympatry region between 
{\em M. m. domesticus} and {\em M. spretus} and its vicinity.
To help maximize genetic differences as part of the design
goals of our pipeline, we selected two publicly available
{\em M. m. domesticus} samples that originated
as far from the sympatry region as possible.
These {\em M. m. domesticus} samples came
from Cyprus.
We also included an {\em M. spretus} sample
from the sympatry region between 
{\em M. m. domesticus} and {\em M. spretus}.
Further, for our controls, we included three wild
{\em M. m. musculus} samples from China as well as
three mice from the C57BL/6J strain.
Genomic coordinates and annotation in our study were based on the 
GRCm38.p2 reference {\em M. musculus} genome (GenBank accession GCA\_000001635.4).

Sequence data for {\em M. musculus} samples were collected
using the Mouse Diversity Array \cite{Yang2009}, either by previous studies or
this study.
Sequence data for all {\em M. musculus} samples other than the German-Hamm
samples were taken from past studies of {\em M. musculus} genetic variation \cite{Yang2011,Didion2012,Guenet200324}.
We sequenced the German-Hamm samples using the same procedure
used to sequence the other samples in our study \cite{Yang2011,Didion2012}. As
part of the sequencing procedure, we genotyped raw sequencing reads using 
MouseDivGeno version 1.0.4 \cite{Didion2012}, 
following the procedure of \cite{Didion2012}.
 We phased {\em M. musculus} genotypes into haplotypes and imputed bases for missing data using FASTPHASE \cite{Scheet2006629}.
The genotyping and phasing analyses were performed with a larger superset of  
362 samples \cite{Didion2012,Yang2011}. 
 Recently, \cite{Keane2011} published the first whole-genome
sequence for {\em M. spretus} as well as additional whole-genome sequences
for {\em M. musculus}.
SNPs, short indels, and structural variation (SV)
were estimated using comparative analysis \cite{Keane2011,Yalcin2011}.
We used the reported SNP data for our {\em M. spretus} sequence.
Since PhyloNet-HMM focuses on evolutionary processes other
than insertions/deletions and structural variation, 
we filtered out loci 
exhibiting short indels and structural variation.
 For outgroup purposes in our analyses,
we used the {\em Rattus norvegicus} reference genome (version RGSC Rnor\_5.0).
To identify orthologous loci between {\em M. musculus} and {\em R. norvegicus}, 
a whole-genome pairwise alignment between the two reference
genomes was obtained from the UCSC Genome 
Browser website \cite{Meyer01012013}. The whole-genome pairwise alignment
was produced using BLASTZ \cite{Schwartz01012003}.
 In summary, the loci used in our study were taken
from the intersection of the following sources:
\begin{enumerate}
\item Wild {\em M. musculus} loci that were genotyped using 
the Mouse Diversity Array \cite{Yang2009}.
\item {\em M. musculus} and {\em M. spretus} loci that exhibited single
nucleotide polymorphism
in the study of \cite{Keane2011}.
\item {\em M. musculus} and {\em M. spretus} loci that
did not exhibit short indel variation in the study of \cite{Keane2011}.
\item {\em M. musculus} and {\em M. spretus} loci that 
did not exhibit structural variation in the study of \cite{Yalcin2011}.
\item Loci in {\em M. musculus} and {\em M. spretus} that had
an orthologous locus in {\em R. norvegicus}.
\end{enumerate}
A total of 414,376 loci from the autosome and allosome 
were used in our data sets.
In total, we created 42 data sets (Supplementary Table \ref{datasets}). Each data set was constructed
using a different set of four haploid mouse genomes and an additional rat genome for outgroup purposes.
 Each  introgression scan examined a wild {\em M. m. domesticus}
sample and an {\em M. spretus} sample from the region of sympatry 
between the two species. In addition, each data set included two {\em M. m. domesticus} samples
taken as far from the sympatry region as possible, for the purpose
of maximizing genetic contrast in the scanning procedure.
 For validation purposes, we performed control 
scans. We scanned {\em M. musculus} samples that were known to not have introgressed
with {\em M. spretus} populations either due to their inbred laboratory origins
or lack of sympatry.
\begin{table*}[!ht]
\centering
\scriptsize
\caption{
{\bf Mouse samples.} 
We obtained new mouse samples and also used existing mouse samples from
previous studies. 
We also included the rat reference genome sequence (RGSC Rnor\_5.0) for use
as an outgroup.
}\label{samples}
\vspace{10mm}
\begin{tabular}{cccccc}
\hline
{\bf Sample name} & {\bf Species/subspecies} & {\bf Origin} & {\bf Gender} & {\bf Source} & {\bf Alias} \\
\hline
Spain-Arenal & {\it M. m. domesticus} & Arenal, Mallorca Island, Spain & Male & \cite{Yang2011} & MWN1279 \\
Spain-RocadelVall\`{e}s & {\it M. m. domesticus} & Roca del Vall\`{e}s, Catalunya, Spain & Female & \cite{Yang2011} & MWN1287  \\
Germany-Hamm-A & {\it M. m. domesticus} & Hamm, North Rhine-Westphalia, Germany & Male & This study & N/A \\
Germany-Hamm-B & {\it M. m. domesticus} & Hamm, North Rhine-Westphalia, Germany & Male & This study & N/A \\
Germany-Hamm-C & {\it M. m. domesticus} & Hamm, North Rhine-Westphalia, Germany & Male & This study & N/A \\
Germany-Hamm-D & {\it M. m. domesticus} & Hamm, North Rhine-Westphalia, Germany & Male & This study & N/A \\
Germany-Hamm-E & {\it M. m. domesticus} & Hamm, North Rhine-Westphalia, Germany & Male & This study & N/A \\
Germany-Hamm-F & {\it M. m. domesticus} & Hamm, North Rhine-Westphalia, Germany & Male & This study & N/A \\
Germany-T\"{u}bingen & {\it M. m. domesticus} & T\"{u}bingen, Germany & Female & \cite{Yang2011} & RDS12763 \\
Germany-Remderoda & {\it M. m. domesticus} & Remderoda, Germany & Male & \cite{Yang2011} & KCT222 \\
Greece-Korinthos & {\it M. m. domesticus} & Korinthos, Velo, Peleponissos, Greece & Male & \cite{Yang2011} & MWN1194 \\
Greece-Laganas & {\it M. m. domesticus} & Laganas, Zakinthos Island, Greece & Male & \cite{Yang2011} & MWN1198 \\
Italy-Menconico & {\it M. m. domesticus} & Menconico, Staffora Valley, Lombardia, Italy & Female & \cite{Yang2011} & MWN1030 \\
Italy-SanGirogio & {\it M. m. domesticus} & San Girogio, Curone Valley, Piamonte, Italy & Male & \cite{Yang2011} & MWN1026 \\
Italy-Cassino & {\it M. m. domesticus} & Cassino, Lazio, Italy & Female & \cite{Yang2011} & MWN1106 \\
Italy-Milazzo & {\it M. m. domesticus} & Milazzo, Olivarella, Sicily, Italy & Female & \cite{Yang2011} & MWN1214 \\
Tunisia-Monastir-A & {\it M. m. domesticus} & Monastir, Tunisia & Male & \cite{Yang2011,Guenet200324} & 22MO \\
Tunisia-Monastir-B & {\it M. m. domesticus} & Monastir, Tunisia & Male & \cite{Yang2011,Guenet200324} & WMP \\
Morocco & {\it M. m. domesticus} & Azemmour, Morocco & Male & \cite{Yang2011,Guenet200324} & DMZ \\
Algeria & {\it M. m. domesticus} & Oran, Algeria & Male & \cite{Yang2011,Guenet200324} & BZO \\
A-background & {\it M. m. domesticus} & Akotiri, Cyprus & Male & \cite{Yang2011,Guenet200324} & DCA \\
B-background & {\it M. m. domesticus} & Paphos, Cyprus & Male & \cite{Yang2011,Guenet200324} & DCP \\
Spretus & {\it M. spretus} & Puerto Real, Cadiz Province, Spain & Male & \cite{Yang2011,Guenet200324} & SPRET/EiJ \\
A-reference & {\it M. m. domesticus} & Classical strain & Female & \cite{Didion2012} & C57BL/6J \\
B-reference & {\it M. m. domesticus} & Classical strain & Female & \cite{Didion2012} & C57BL/6J \\
C-reference & {\it M. m. domesticus} & Classical strain & Female & \cite{Didion2012} & C57BL/6J \\
A-musculus & {\it M. m. musculus} & Urumqi, Xinjiang, China & Male & \cite{Yang2011} & Yu2097m \\
B-musculus & {\it M. m. musculus} & Hebukesaier, Xinjiang, China & Female & \cite{Yang2011} & Yu2120f \\
C-musculus & {\it M. m. musculus} & Yutian, Xinjiang, China & Male & \cite{Yang2011} & Yu2115m \\
\hline
\end{tabular}
\end{table*}

\clearpage
\begin{table*}[!ht]
\centering
\scriptsize
\caption{
{\bf Datasets used in our study.} 
Each dataset consisted of four {\it Mus} genomes
and 
an additional rat genome (the rat reference genome sequenc RGSC Rnor\_5.0)
for outgroup purposes.
}\label{datasets}
\vspace{10mm}
\begin{tabular}{cc}
\hline
{\bf Scan type} & {\bf Set of {\it Mus} samples} \\
\hline
Introgression & Spain-Arenal haplotype 1, A-background haplotype 1, B-background haplotype 1, Spretus haplotype 1 \\
Introgression & Spain-Arenal haplotype 2, A-background haplotype 1, B-background haplotype 1, Spretus haplotype 1 \\
Introgression & Spain-RocadelVall\`{e}s haplotype 1, A-background haplotype 1, B-background haplotype 1, Spretus haplotype 1 \\
Introgression & Spain-RocadelVall\`{e}s haplotype 2, A-background haplotype 1, B-background haplotype 1, Spretus haplotype 1 \\
Introgression & Germany-Hamm-A haplotype 1, A-background haplotype 1, B-background haplotype 1, Spretus haplotype 1 \\
Introgression & Germany-Hamm-A haplotype 2, A-background haplotype 1, B-background haplotype 1, Spretus haplotype 1 \\
Introgression & Germany-Hamm-B haplotype 1, A-background haplotype 1, B-background haplotype 1, Spretus haplotype 1 \\
Introgression & Germany-Hamm-B haplotype 2, A-background haplotype 1, B-background haplotype 1, Spretus haplotype 1 \\
Introgression & Germany-Hamm-C haplotype 1, A-background haplotype 1, B-background haplotype 1, Spretus haplotype 1 \\
Introgression & Germany-Hamm-C haplotype 2, A-background haplotype 1, B-background haplotype 1, Spretus haplotype 1 \\
Introgression & Germany-Hamm-D haplotype 1, A-background haplotype 1, B-background haplotype 1, Spretus haplotype 1 \\
Introgression & Germany-Hamm-D haplotype 2, A-background haplotype 1, B-background haplotype 1, Spretus haplotype 1 \\
Introgression & Germany-Hamm-E haplotype 1, A-background haplotype 1, B-background haplotype 1, Spretus haplotype 1 \\
Introgression & Germany-Hamm-E haplotype 2, A-background haplotype 1, B-background haplotype 1, Spretus haplotype 1 \\
Introgression & Germany-Hamm-F haplotype 1, A-background haplotype 1, B-background haplotype 1, Spretus haplotype 1 \\
Introgression & Germany-Hamm-F haplotype 2, A-background haplotype 1, B-background haplotype 1, Spretus haplotype 1 \\
Introgression & Germany-T\"{u}bingen haplotype 1, A-background haplotype 1, B-background haplotype 1, Spretus haplotype 1 \\
Introgression & Germany-T\"{u}bingen haplotype 2, A-background haplotype 1, B-background haplotype 1, Spretus haplotype 1 \\
Introgression & Germany-Remderoda haplotype 1, A-background haplotype 1, B-background haplotype 1, Spretus haplotype 1 \\
Introgression & Germany-Remderoda haplotype 2, A-background haplotype 1, B-background haplotype 1, Spretus haplotype 1 \\
Introgression & Greece-Korinthos haplotype 1, A-background haplotype 1, B-background haplotype 1, Spretus haplotype 1 \\
Introgression & Greece-Korinthos haplotype 2, A-background haplotype 1, B-background haplotype 1, Spretus haplotype 1 \\
Introgression & Greece-Laganas haplotype 1, A-background haplotype 1, B-background haplotype 1, Spretus haplotype 1 \\
Introgression & Greece-Laganas haplotype 2, A-background haplotype 1, B-background haplotype 1, Spretus haplotype 1 \\
Introgression & Italy-Menconico haplotype 1, A-background haplotype 1, B-background haplotype 1, Spretus haplotype 1 \\
Introgression & Italy-Menconico haplotype 2, A-background haplotype 1, B-background haplotype 1, Spretus haplotype 1 \\
Introgression & Italy-SanGirogio haplotype 1, A-background haplotype 1, B-background haplotype 1, Spretus haplotype 1 \\
Introgression & Italy-SanGirogio haplotype 2, A-background haplotype 1, B-background haplotype 1, Spretus haplotype 1 \\
Introgression & Italy-Cassino haplotype 1, A-background haplotype 1, B-background haplotype 1, Spretus haplotype 1 \\
Introgression & Italy-Cassino haplotype 2, A-background haplotype 1, B-background haplotype 1, Spretus haplotype 1 \\
Introgression & Italy-Milazzo haplotype 1, A-background haplotype 1, B-background haplotype 1, Spretus haplotype 1 \\
Introgression & Italy-Milazzo haplotype 2, A-background haplotype 1, B-background haplotype 1, Spretus haplotype 1 \\
Introgression & Tunisia-Monastir-A haplotype 1, A-background haplotype 1, B-background haplotype 1, Spretus haplotype 1 \\
Introgression & Tunisia-Monastir-A haplotype 2, A-background haplotype 1, B-background haplotype 1, Spretus haplotype 1 \\
Introgression & Tunisia-Monastir-B haplotype 1, A-background haplotype 1, B-background haplotype 1, Spretus haplotype 1 \\
Introgression & Tunisia-Monastir-B haplotype 2, A-background haplotype 1, B-background haplotype 1, Spretus haplotype 1 \\
Introgression & Morocco haplotype 1, A-background haplotype 1, B-background haplotype 1, Spretus haplotype 1 \\
Introgression & Morocco haplotype 2, A-background haplotype 1, B-background haplotype 1, Spretus haplotype 1 \\
Introgression & Algeria haplotype 1, A-background haplotype 1, B-background haplotype 1, Spretus haplotype 1 \\
Introgression & Algeria haplotype 2, A-background haplotype 1, B-background haplotype 1, Spretus haplotype 1 \\
Control & A-reference haplotype 1, B-reference haplotype 1, C-reference haplotype 1, Spretus haplotype 1 \\
Control & A-musculus haplotype 1, B-musculus haplotype 1, C-musculus haplotype 1, Spretus haplotype 1 \\
\hline
\end{tabular}
\end{table*}


\subsection{Introgression detection}
We used PhyloNet-HMM \cite{Liu2014}, implemented as part of the 
PhyloNet software package \cite{Than2008}, to scan
{\em M. musculus} genomes for 
segments with introgressed origin from {\em M. spretus}.
A strength of this approach is
its ability to search for introgression while also accounting for
incomplete lineage sorting, point mutations, recombination,
and dependence across loci.
 In brief, PhyloNet-HMM 
examines local patterns of phylogenetic incongruence
under a statistical model that combines the basic coalescent model \cite{Kingman1982}
with a hidden Markov model (HMM) \cite{Rabiner89atutorial} approach.
Introgression, recombination, and incomplete lineage sorting
are captured by switching between HMM states.
The substitution process is modeled by state emissions.
Dependence among neighboring loci in the genome is captured by the Markov property
of the HMM.
PhyloNet-HMM works
directly from primary biomolecular sequence data and
incorporates finite sites models of nucleotide substitution.
Therefore, PhyloNet-HMM is capable of
analyzing divergent sequences that may have undergone multiple
substitutions at each locus (as in the case of the lineages between
mouse, rat, and their most recent common ancestor). 

In our study, each diploid wild {\it M. m. domesticus} sample under
study (20 in total) was represented by two datasets ${\cal G}_{h} = \{O_{c,h}\}$ where $h \in \{1, 2\}$, one dataset for
each of the sample's two haploid genome sequences. Each dataset ${\cal G}_{h}$ consisted of
$1 \leq c \leq 20$ haploid chromosomal alignments $O_{c,h}$. 

\begin{figure}[ht]
\begin{center}
\centerline{\includegraphics[scale=0.4]{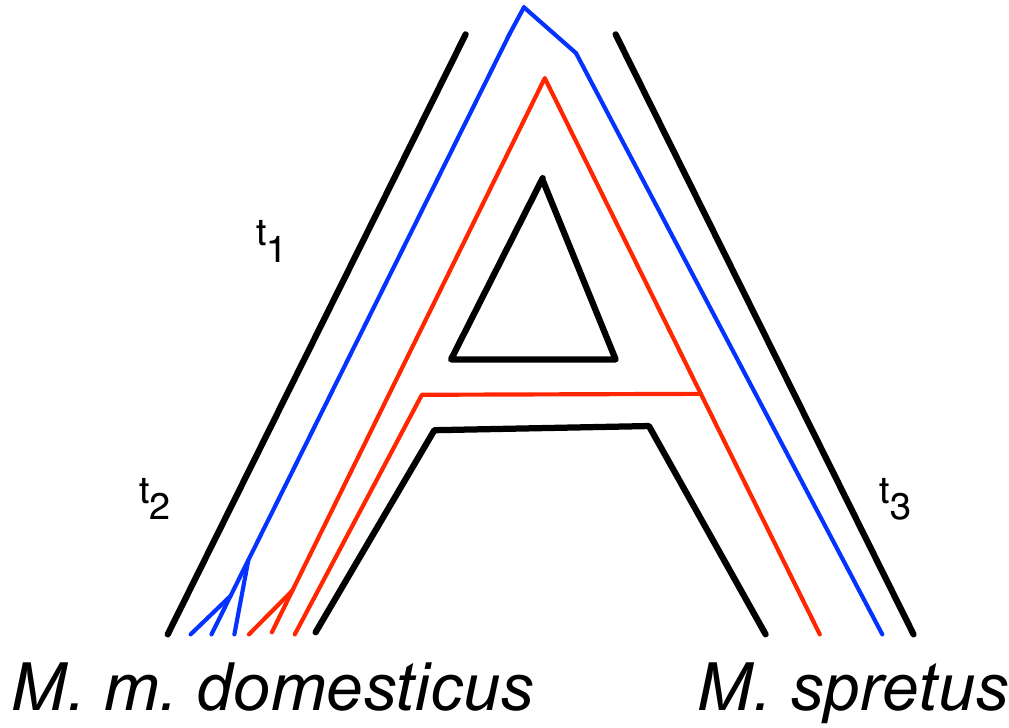}}
\end{center}
\caption{
{\bf A phylogeny of divergence followed by hybridization between 
{\em M. m. domesticus} and {\em M. spretus}.} 
Branches of the phylogeny have lengths $t_1$, $t_2$, and $t_3$. 
PhyloNet-HMM infers parameter values corresponding
to the branch lengths, 
alongside inference of other parameters
in its combined model. (Inference for $t_3$ would
require at least one additional {\it M. spretus} sample in our data sets.)
The incongruence between the two local trees shown here is due to hybridization and 
incomplete lineage sorting (ILS). 
}\label{phylogenies-for-model-selection}
\end{figure}

The species network model utilized by PhyloNet-HMM for this study
is shown in Fig. \ref{phylogenies-for-model-selection}.

PhyloNet-HMM was run using a partitioned approach 
where the set of model parameters consisted of
genome-wide time parameters $t_1, t_2, t_3$ 
and chromosome-specific state-switching parameters $\delta_c$
for $1 \leq c \leq 20$. The analysis
proceeded in two phases. 
First, the 
times $t_1, t_2, t_3$ 
were estimated using chromosome 7, which contains the {\it Vkorc1} gene for which recent introgression was described in prior literature,
from the Spain-RocadelVall\`{e}s dataset, which utilizes samples
from the region of sympatry between {\it M. m. domesticus} and {\it M. spretus}.
The times were estimated as  

\[
\argmax_{t_1, t_2, t_3, \delta_7} \prod_{h \in \{1,2\}} {\bf P}(O_{7,h} | t_1, t_2, t_3, \delta_7).
\]

Next, using the estimates $t_1', t_2', t_3'$ for the model parameters
$t_1, t_2, t_3$, we ran PhyloNet-HMM to analyze 
all 20 chromosomes across the 20 wild {\it M. m. domesticus} samples under study
(400 chromosomes in total). For each chromosome $c$, 
the optimization procedure involved maximization of the
criterion

\[
\max_{\delta_c} \prod_{h \in \{1,2\}} {\bf P}(O_{c,h} | t_1', t_2', t_3', \delta_c).
\]





\subsection{Selective sweep scans}
We used XP-CLR \cite{Chen01032010} version 1.0 to scan for selective sweep patterns.
Both methods were run using their default settings.
The samples used for these scans consisted of subsets of the 362 superset of samples used 
for genotyping/phasing: the wild {\em M. m. domesticus} and wild {\em M. m. musculus}
samples from \cite{Yang2011} and six additional wild {\em M. m. domesticus} 
samples from Hamm, North Rhine-Westphalia, Germany.


\subsection{Functional analysis}
We used BiNGO \cite{Maere15082005} version 2.44 to identify
significantly enriched Gene Ontology (GO) \cite{Ashburner2000} terms in introgressed genes
(hypergeometric test with Benjamini-Hochberg \cite{BenjaminiHochberg} correction; $\alpha < 0.05$).
The GO terms were retrieved from \url{www.geneontology.org/GO.downloads.ontology.shtml}
with version number 1.2 and dated October 31, 2013.
Gene associations with GO terms were 
retrieved from \url{ftp.informatics.jax.org/pub/reports/index.html#go} with
version number 1.3 and dated November 1, 2013.
We used custom scripts to convert the GO term database and gene association file
into formats suitable for input to BiNGO.

\end{article}

\begin{figure*}[!ht]
\begin{center}
\centerline{\includegraphics[scale=1.7]{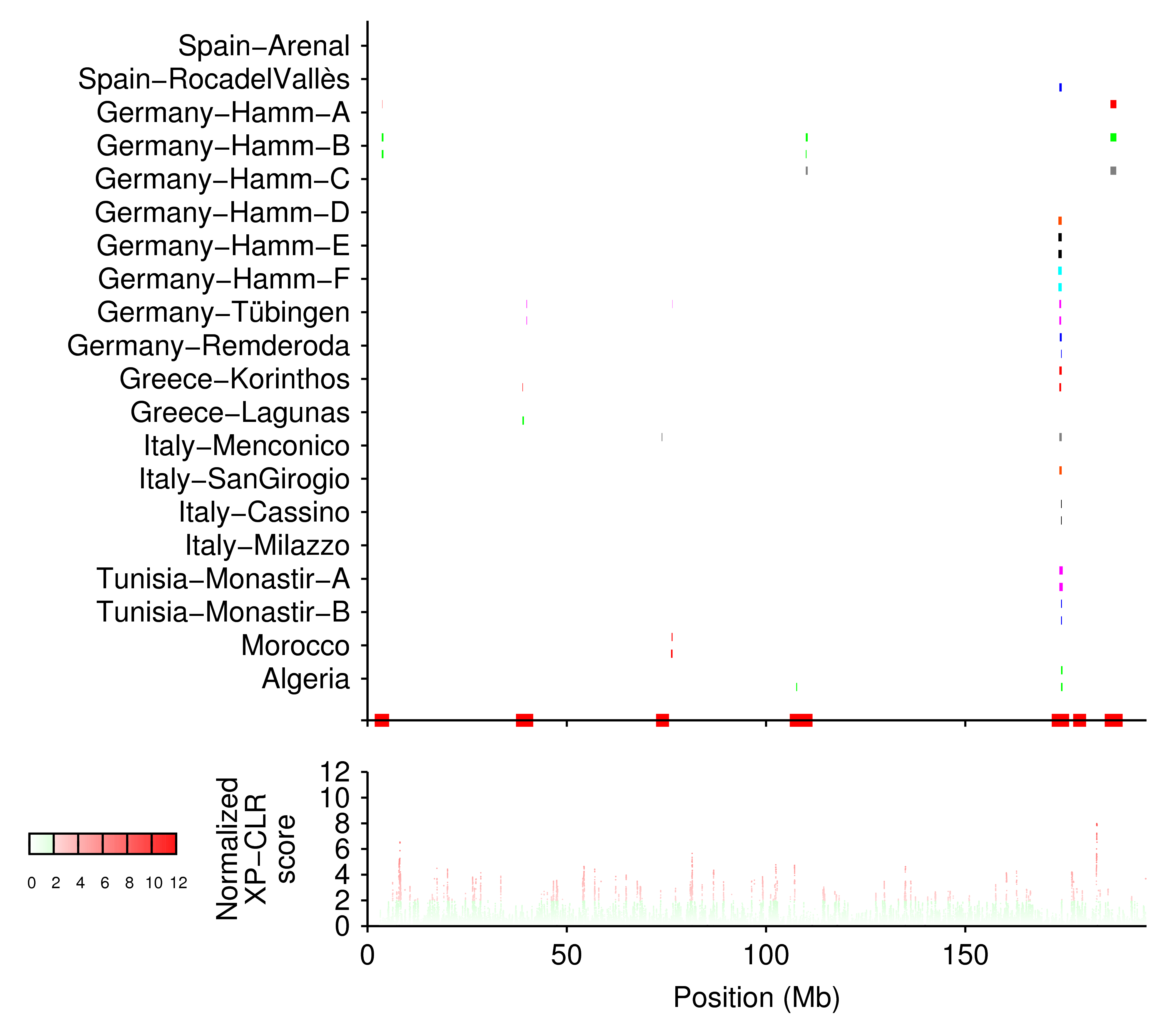}}
\end{center}
\caption{
{\bf Introgression scans of chromosome 1.} 
PhyloNet-HMM was used to detect introgression based upon
a minimum posterior decoding probability of 95\% (see Materials and Methods).
(Top panel) Introgressed genomic tracts are shown for the twenty diploid {\it M. m. domesticus} samples
under analysis in our study. Each sample is shown in a separate track (with adjacent samples colored using different colors to aid visibility). 
Within a track, 
the two haploid genomes for each diploid sample are adjacent.
Along the horizontal axis, the locations of introgressed genes are marked with red squares.
(Bottom panel) Selective sweep statistics
are shown, which are 
normalized XP-CLR scores \cite{Chen01032010} based on a comparison of rodenticide-resistant to 
rodenticide-susceptible wild {\em M. m. domesticus} samples. 
Data points vary in height (low to high corresponding
to scores from $0$ to an upper bound on the values encountered in our study),
opacity (invisible to opaque, similarly),
and color. We used an empirical coloring scheme
to emphasize relative intensity of selective sweep measures, where
data points greater than 2 are colored red, and green otherwise.
}
\label{hybrid-chr1}
\end{figure*}

\clearpage

\begin{figure*}[!ht]
\begin{center}
\centerline{\includegraphics[scale=1.7]{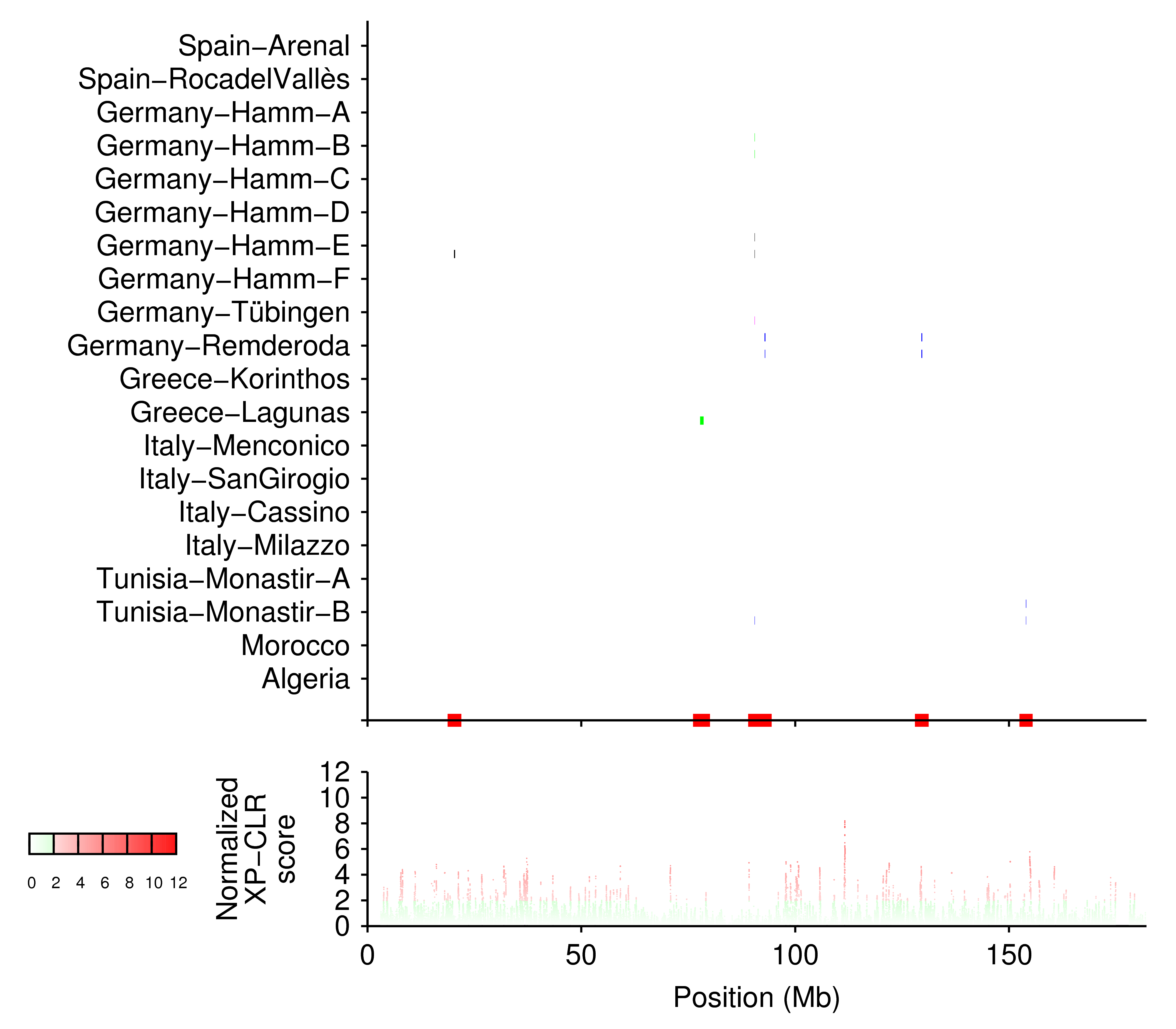}}
\end{center}
\caption{
{\bf Introgression scans of chromosome 2.} 
Figure layout and description are otherwise identical to Figure \ref{hybrid-chr1}.
}
\label{hybrid-chr2}
\end{figure*}

\clearpage

\begin{figure*}[!ht]
\begin{center}
\centerline{\includegraphics[scale=1.7]{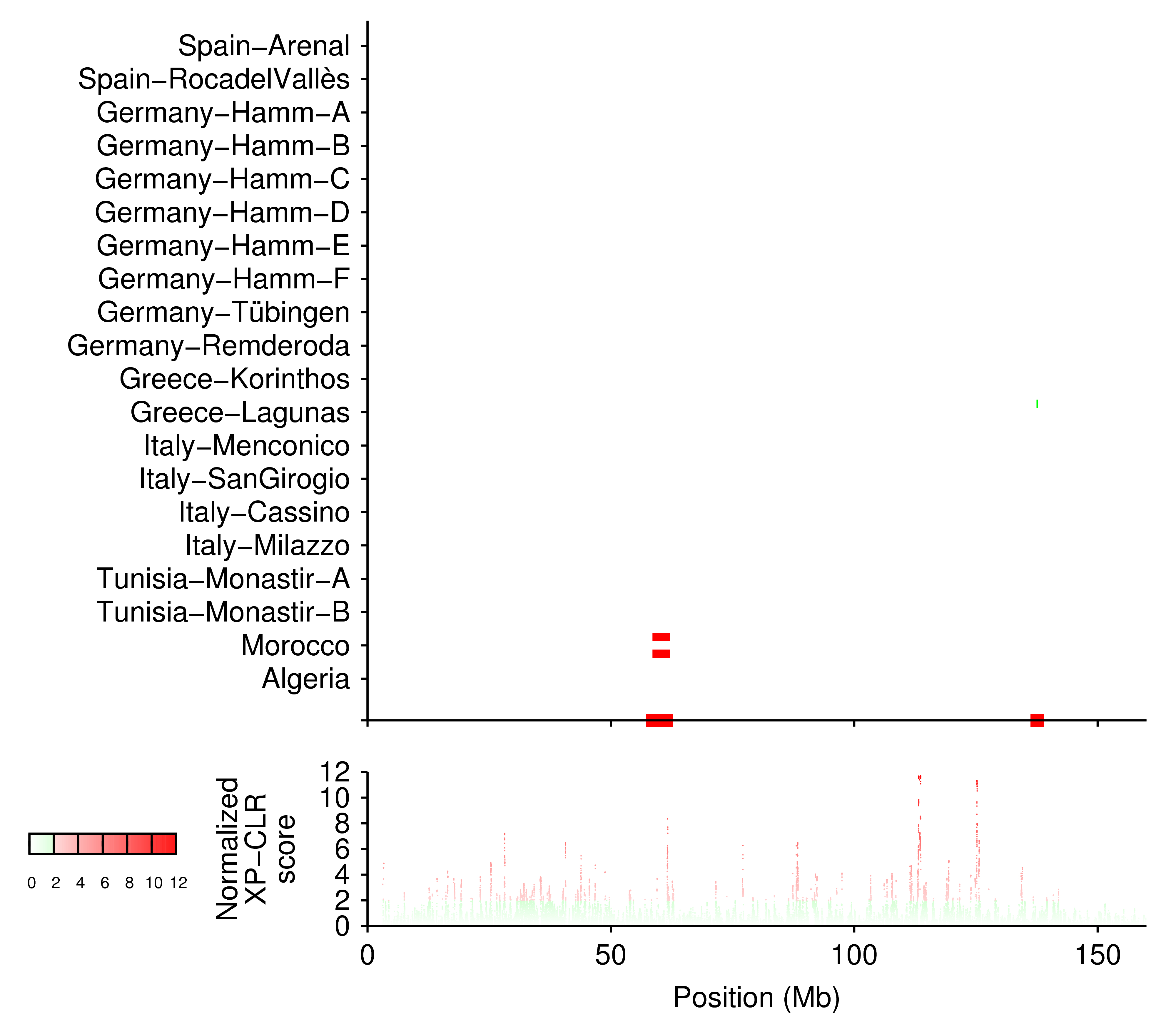}}
\end{center}
\caption{
{\bf Introgression scans of chromosome 3.} 
Figure layout and description are otherwise identical to Figure \ref{hybrid-chr1}.
}
\label{hybrid-chr3}
\end{figure*}

\clearpage

\begin{figure*}[!ht]
\begin{center}
\centerline{\includegraphics[scale=1.7]{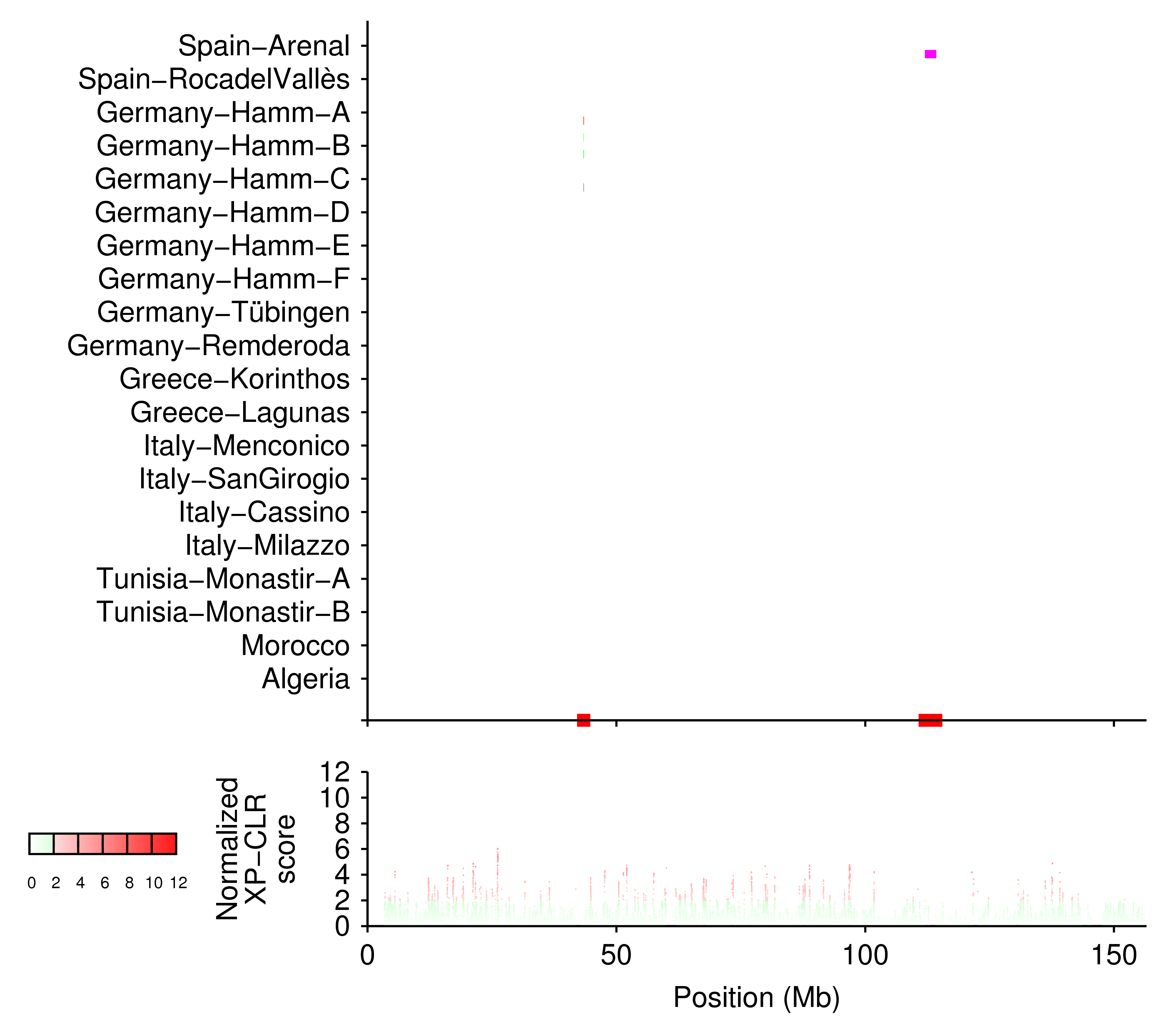}}
\end{center}
\caption{
{\bf Introgression scans of chromosome 4.} 
Figure layout and description are otherwise identical to Figure \ref{hybrid-chr1}.
}
\label{hybrid-chr4}
\end{figure*}

\clearpage

\begin{figure*}[!ht]
\begin{center}
\centerline{\includegraphics[scale=1.7]{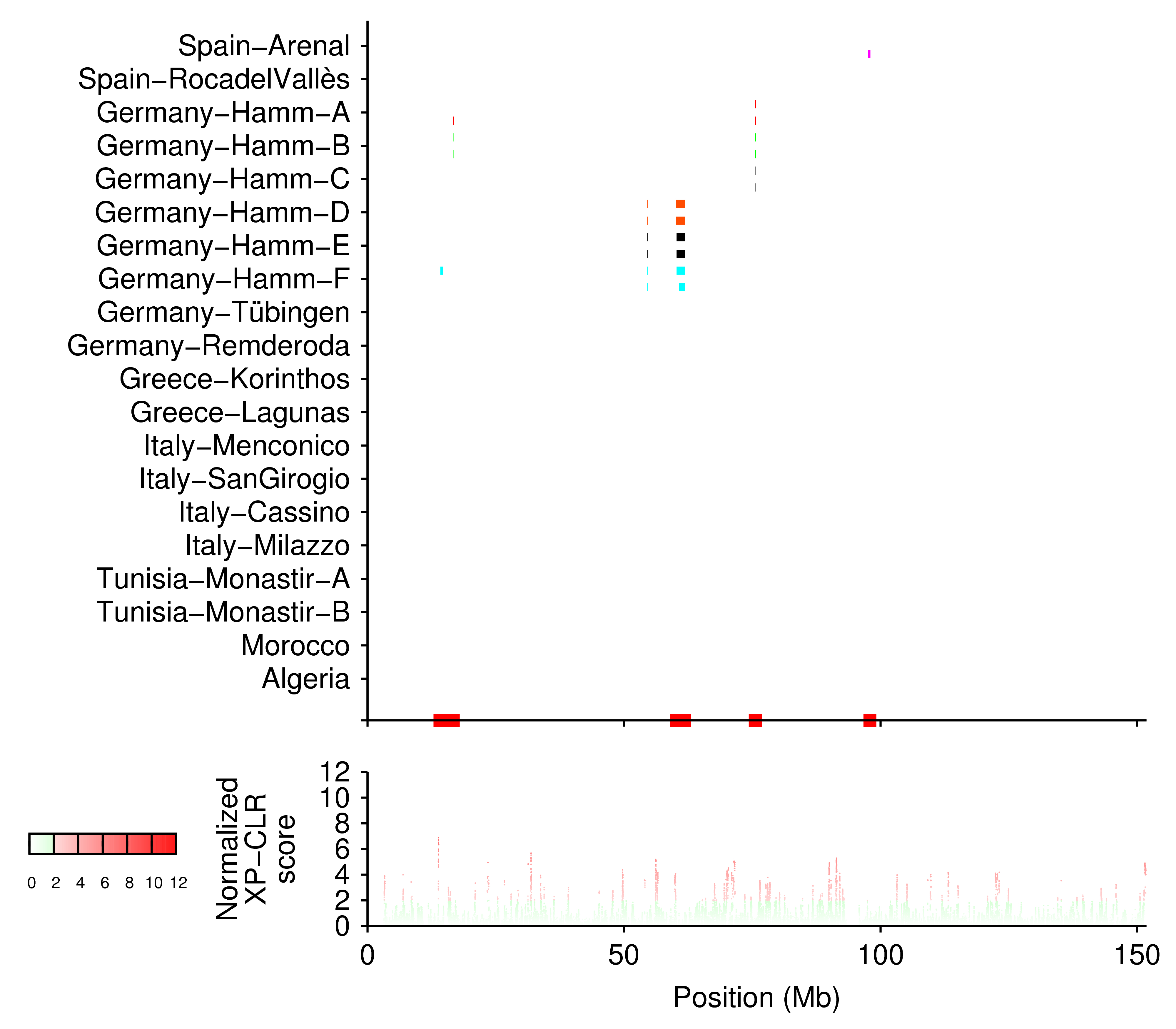}}
\end{center}
\caption{
{\bf Introgression scans of chromosome 5.} 
Figure layout and description are otherwise identical to Figure \ref{hybrid-chr1}.
}
\label{hybrid-chr5}
\end{figure*}

\clearpage

\begin{figure*}[!ht]
\begin{center}
\centerline{\includegraphics[scale=1.7]{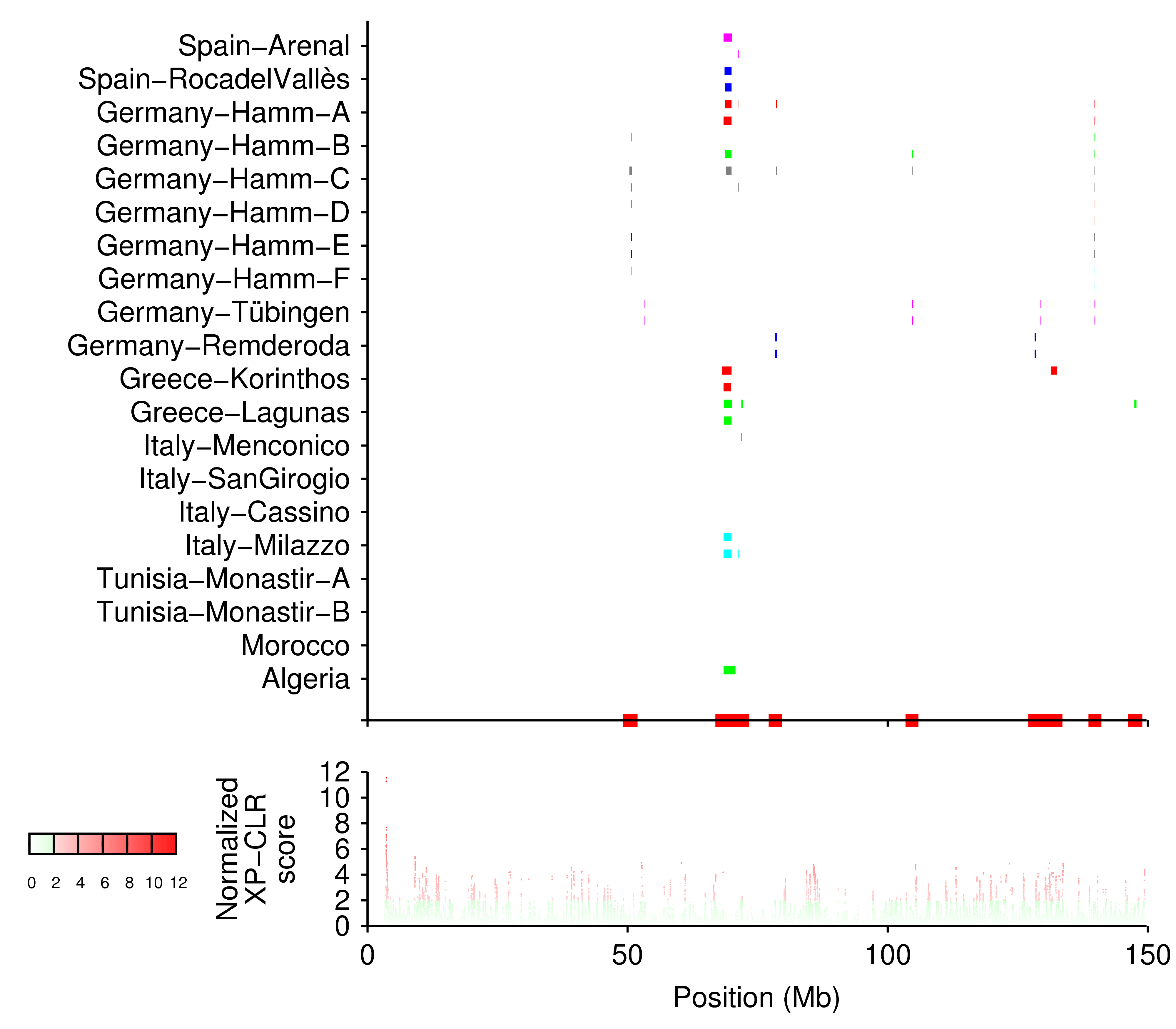}}
\end{center}
\caption{
{\bf Introgression scans of chromosome 6.} 
Figure layout and description are otherwise identical to Figure \ref{hybrid-chr1}.
}
\label{hybrid-chr6}
\end{figure*}

\clearpage

\begin{figure*}[!ht]
\begin{center}
\centerline{\includegraphics[scale=1.7]{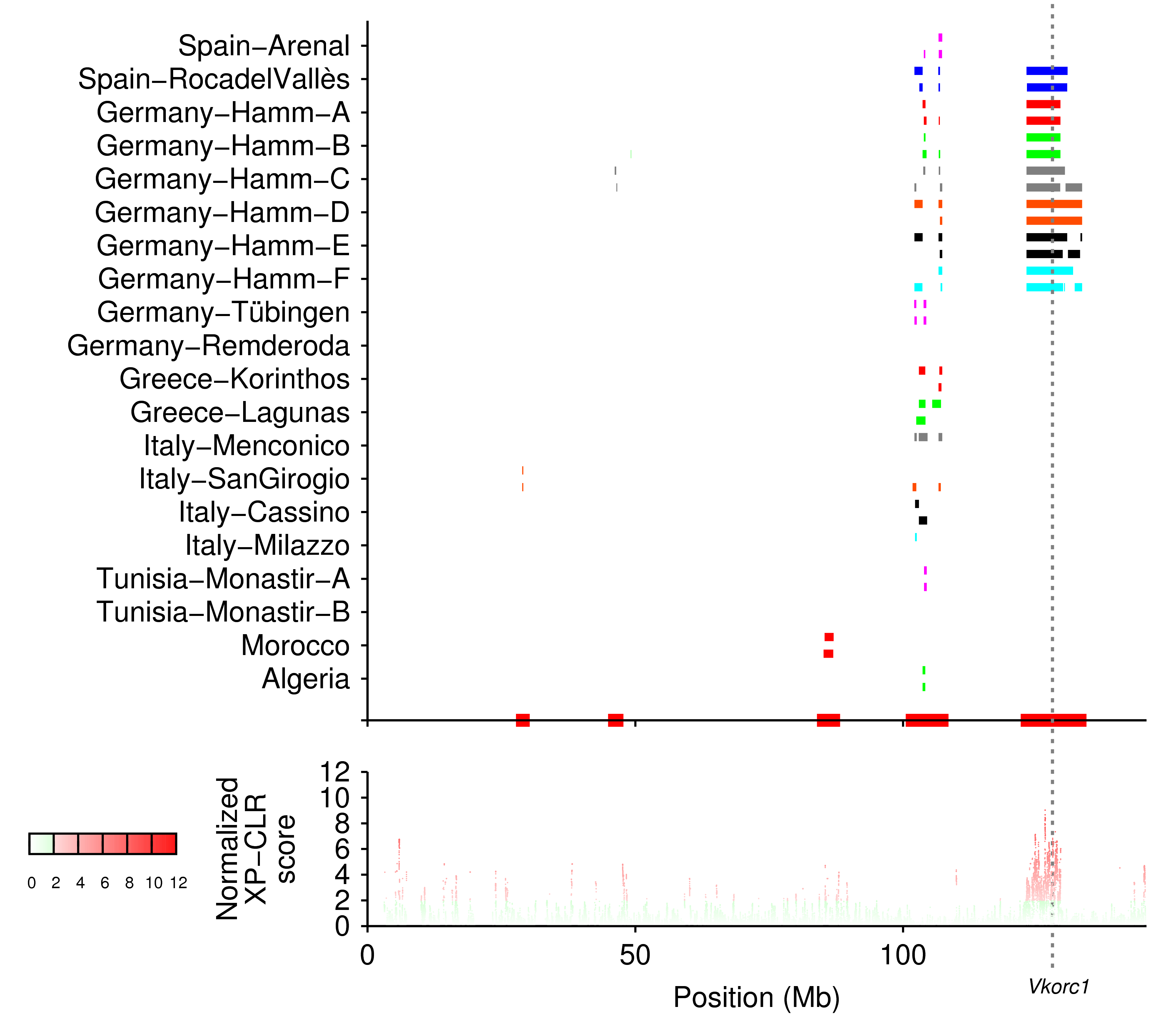}}
\end{center}
\caption{
{\bf Introgression scans of chromosome 7.} 
Figure layout and description are otherwise identical to Figure \ref{hybrid-chr1}.
}
\label{hybrid-chr7}
\end{figure*}

\clearpage

\begin{figure*}[!ht]
\begin{center}
\centerline{\includegraphics[scale=1.7]{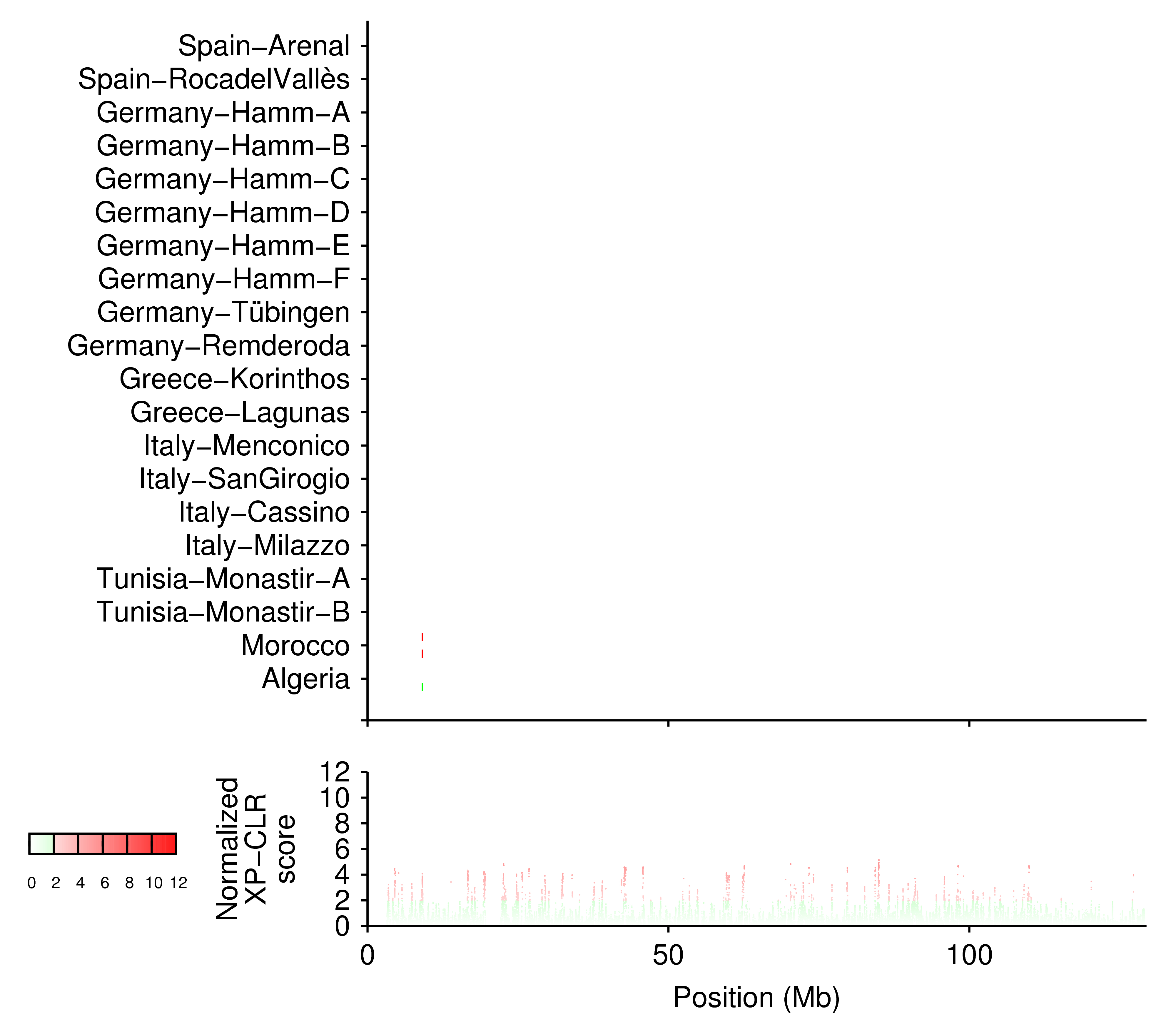}}
\end{center}
\caption{
{\bf Introgression scans of chromosome 8.} 
Figure layout and description are otherwise identical to Figure \ref{hybrid-chr1}.
}
\label{hybrid-chr8}
\end{figure*}

\clearpage

\begin{figure*}[!ht]
\begin{center}
\centerline{\includegraphics[scale=1.7]{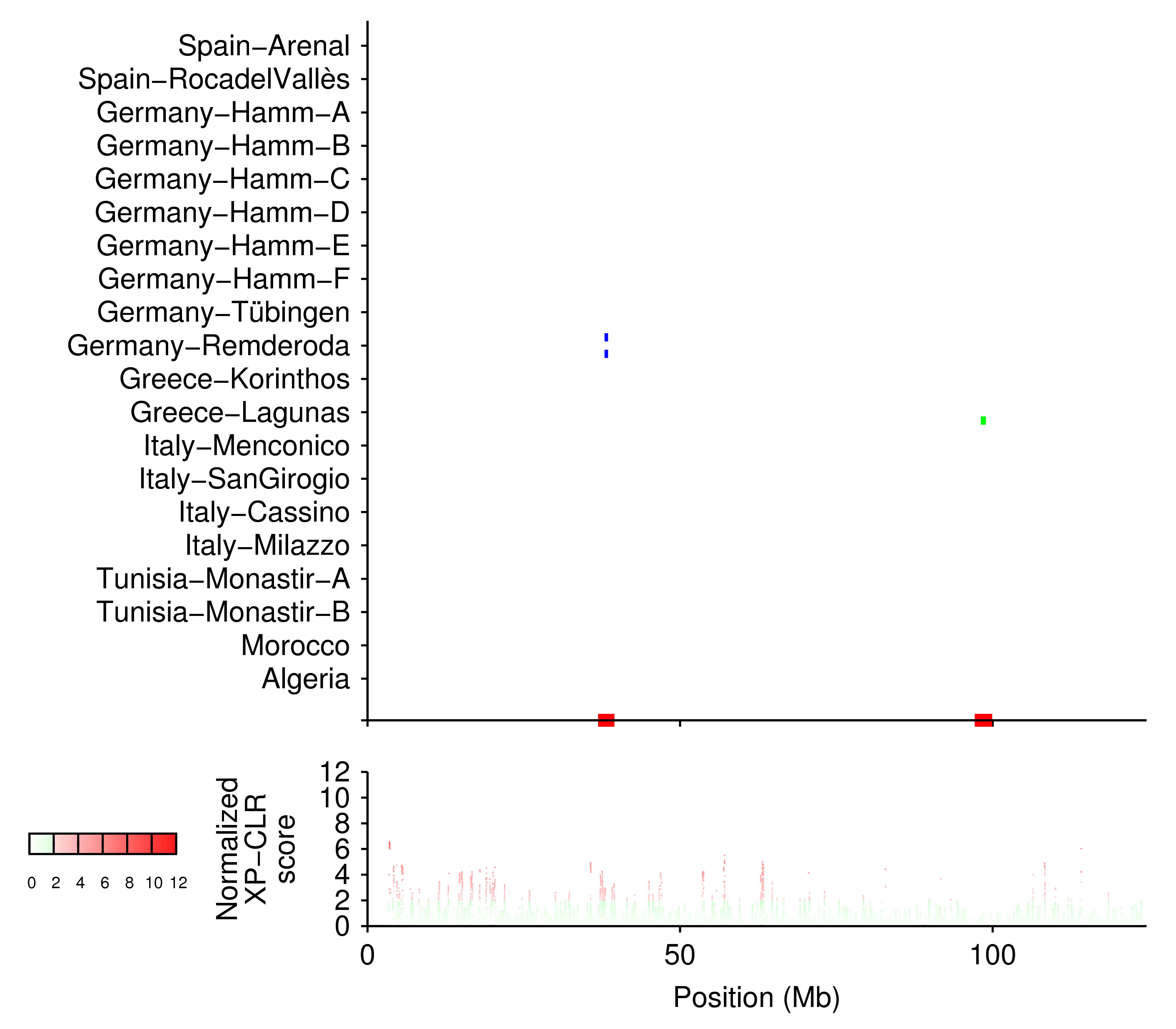}}
\end{center}
\caption{
{\bf Introgression scans of chromosome 9.} 
Figure layout and description are otherwise identical to Figure \ref{hybrid-chr1}.
}
\label{hybrid-chr9}
\end{figure*}

\clearpage

\begin{figure*}[!ht]
\begin{center}
\centerline{\includegraphics[scale=1.7]{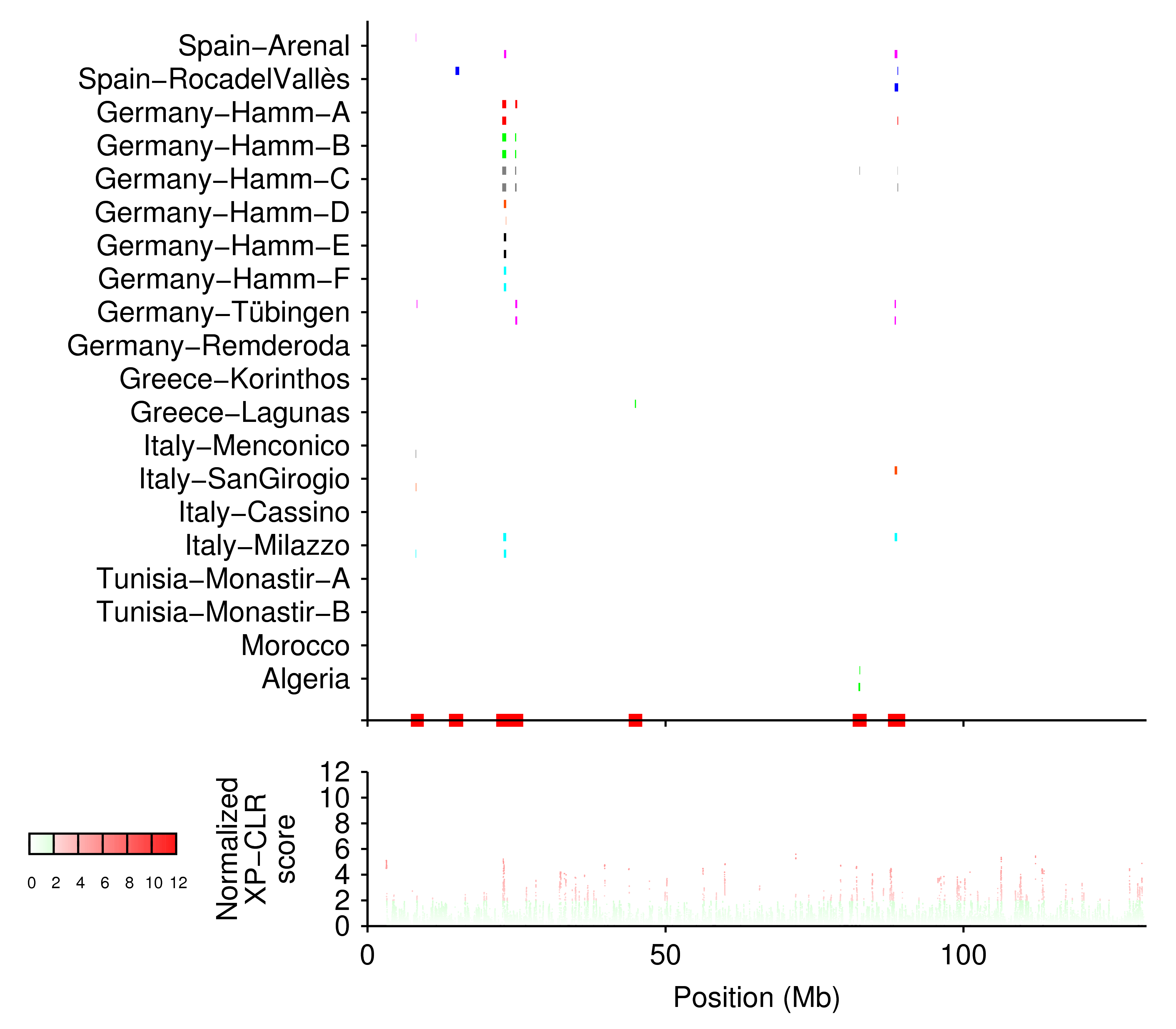}}
\end{center}
\caption{
{\bf Introgression scans of chromosome 10.} 
Figure layout and description are otherwise identical to Figure \ref{hybrid-chr1}.
}
\label{hybrid-chr10}
\end{figure*}

\clearpage

\begin{figure*}[!ht]
\begin{center}
\centerline{\includegraphics[scale=1.7]{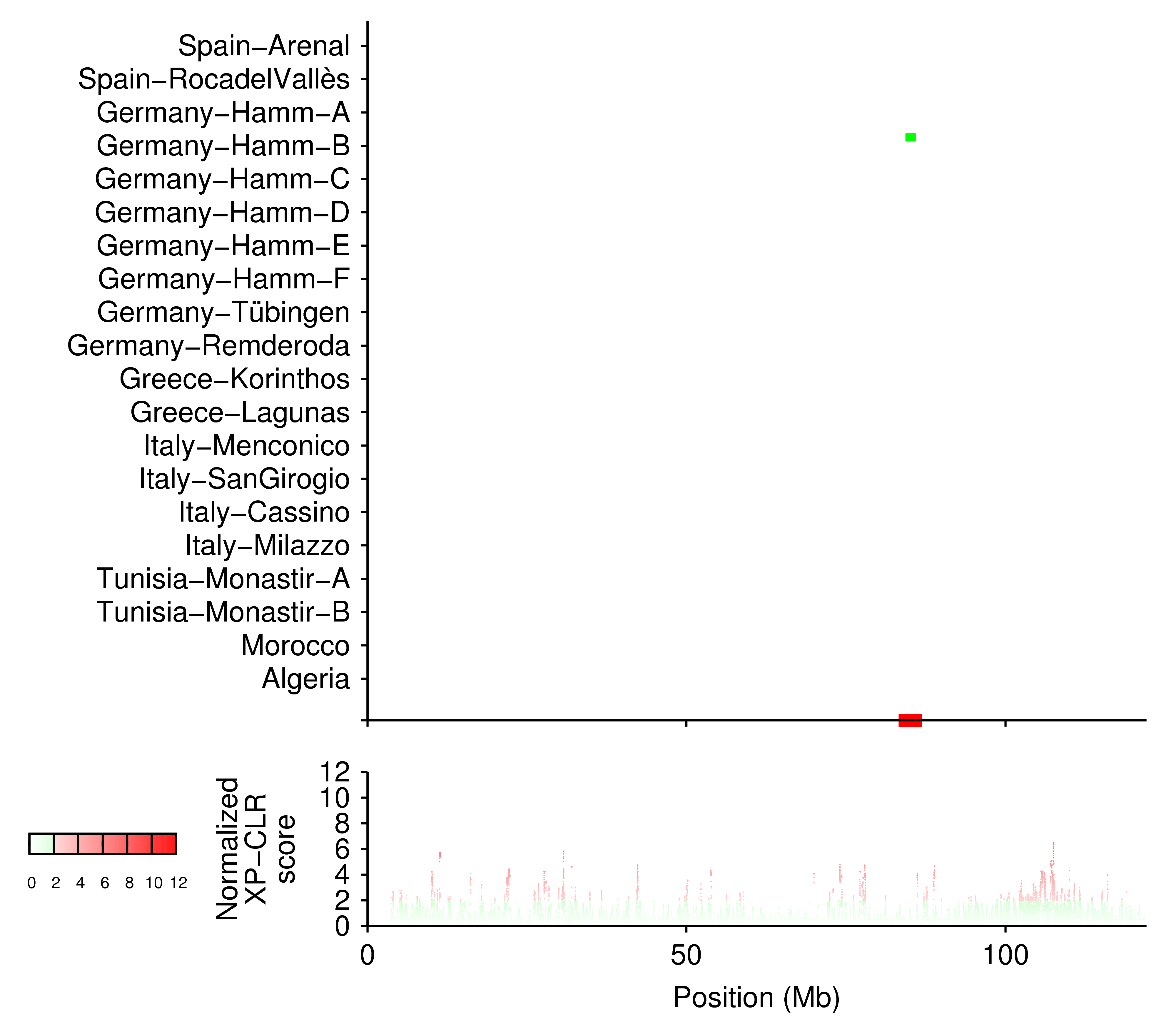}}
\end{center}
\caption{
{\bf Introgression scans of chromosome 11.} 
Figure layout and description are otherwise identical to Figure \ref{hybrid-chr1}.
}
\label{hybrid-chr11}
\end{figure*}

\clearpage

\begin{figure*}[!ht]
\begin{center}
\centerline{\includegraphics[scale=1.7]{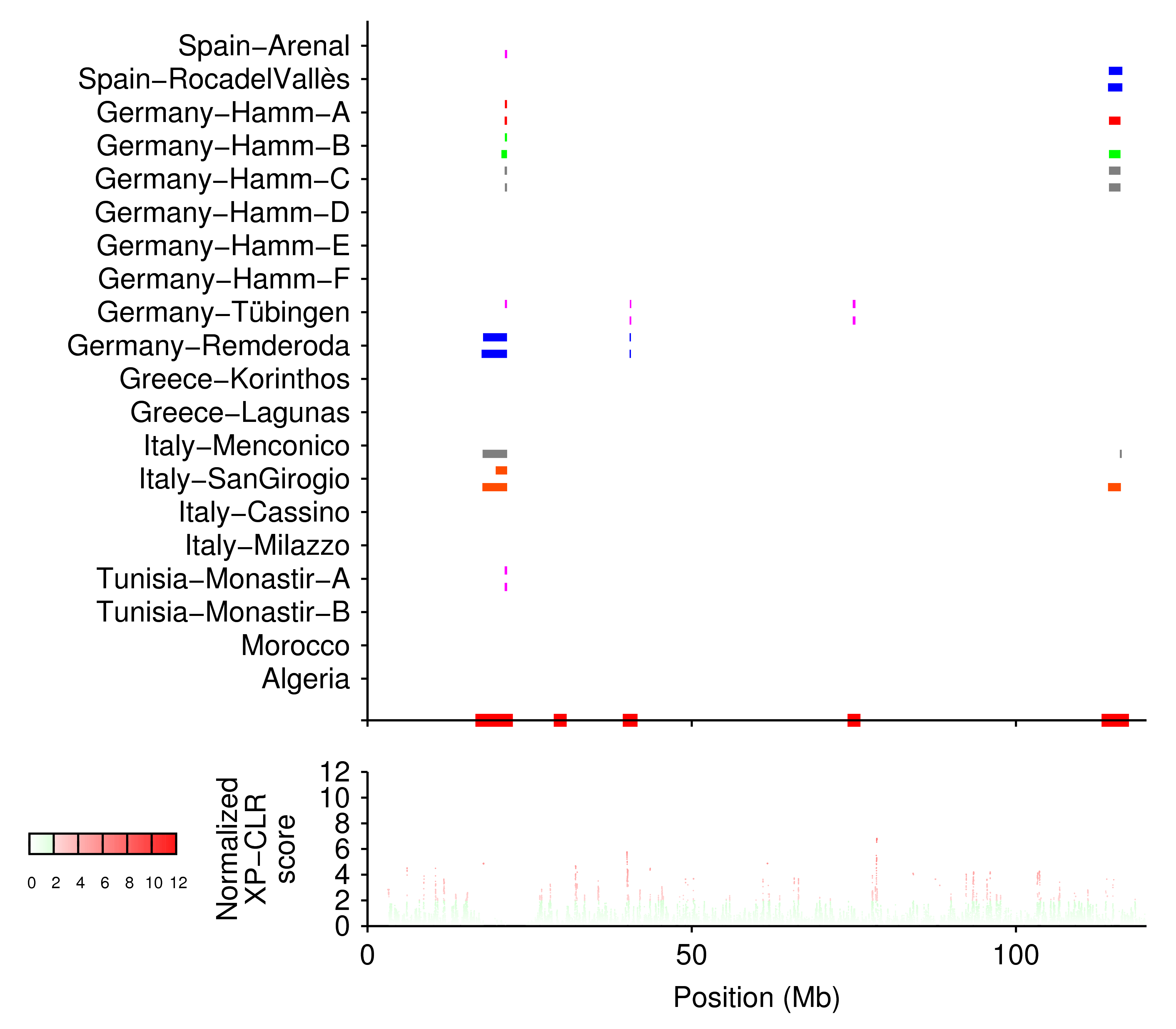}}
\end{center}
\caption{
{\bf Introgression scans of chromosome 12.} 
Figure layout and description are otherwise identical to Figure \ref{hybrid-chr1}.
}
\label{hybrid-chr12}
\end{figure*}

\clearpage

\begin{figure*}[!ht]
\begin{center}
\centerline{\includegraphics[scale=1.7]{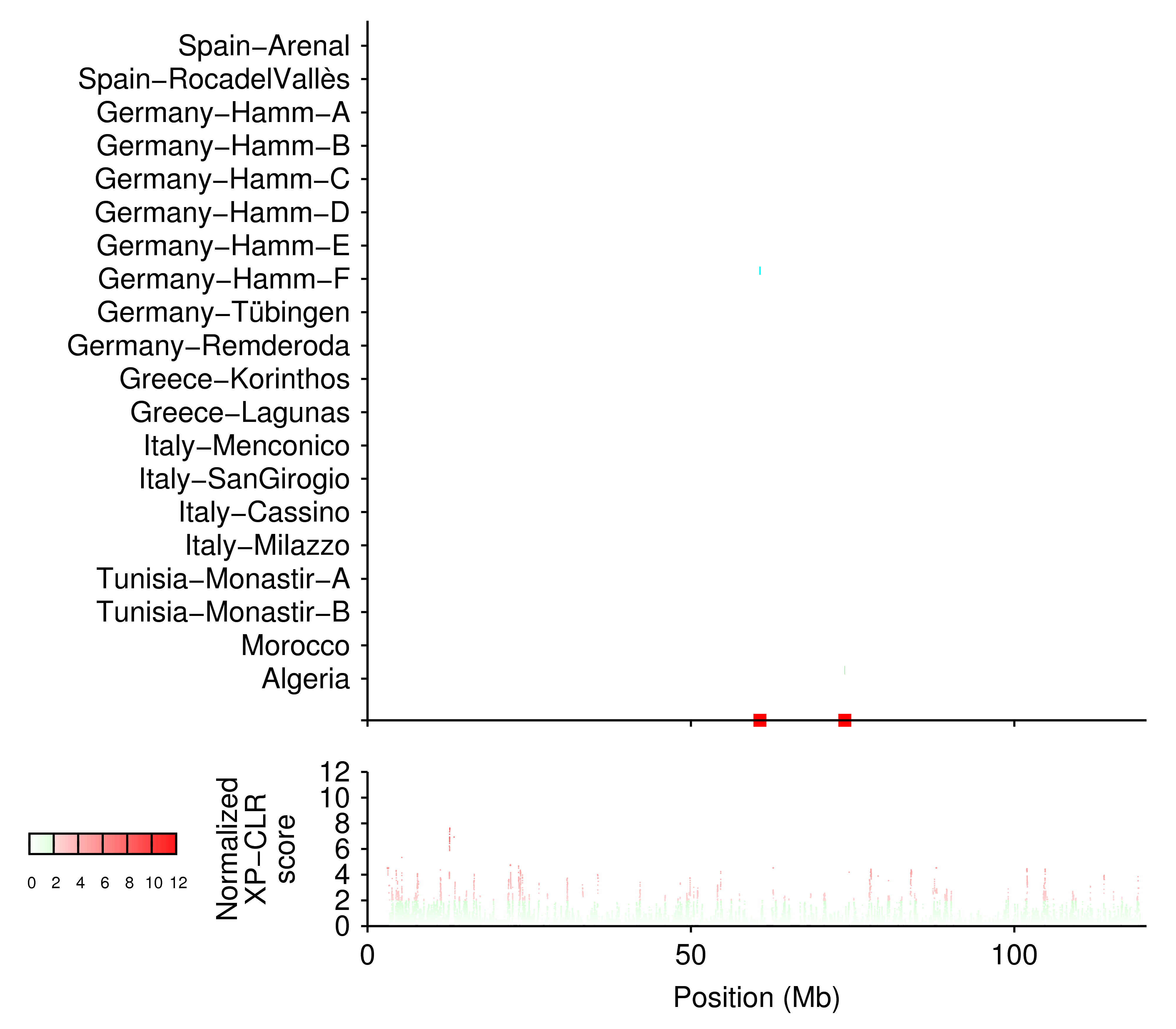}}
\end{center}
\caption{
{\bf Introgression scans of chromosome 13.} 
Figure layout and description are otherwise identical to Figure \ref{hybrid-chr1}.
}
\label{hybrid-chr13}
\end{figure*}

\clearpage

\begin{figure*}[!ht]
\begin{center}
\centerline{\includegraphics[scale=1.7]{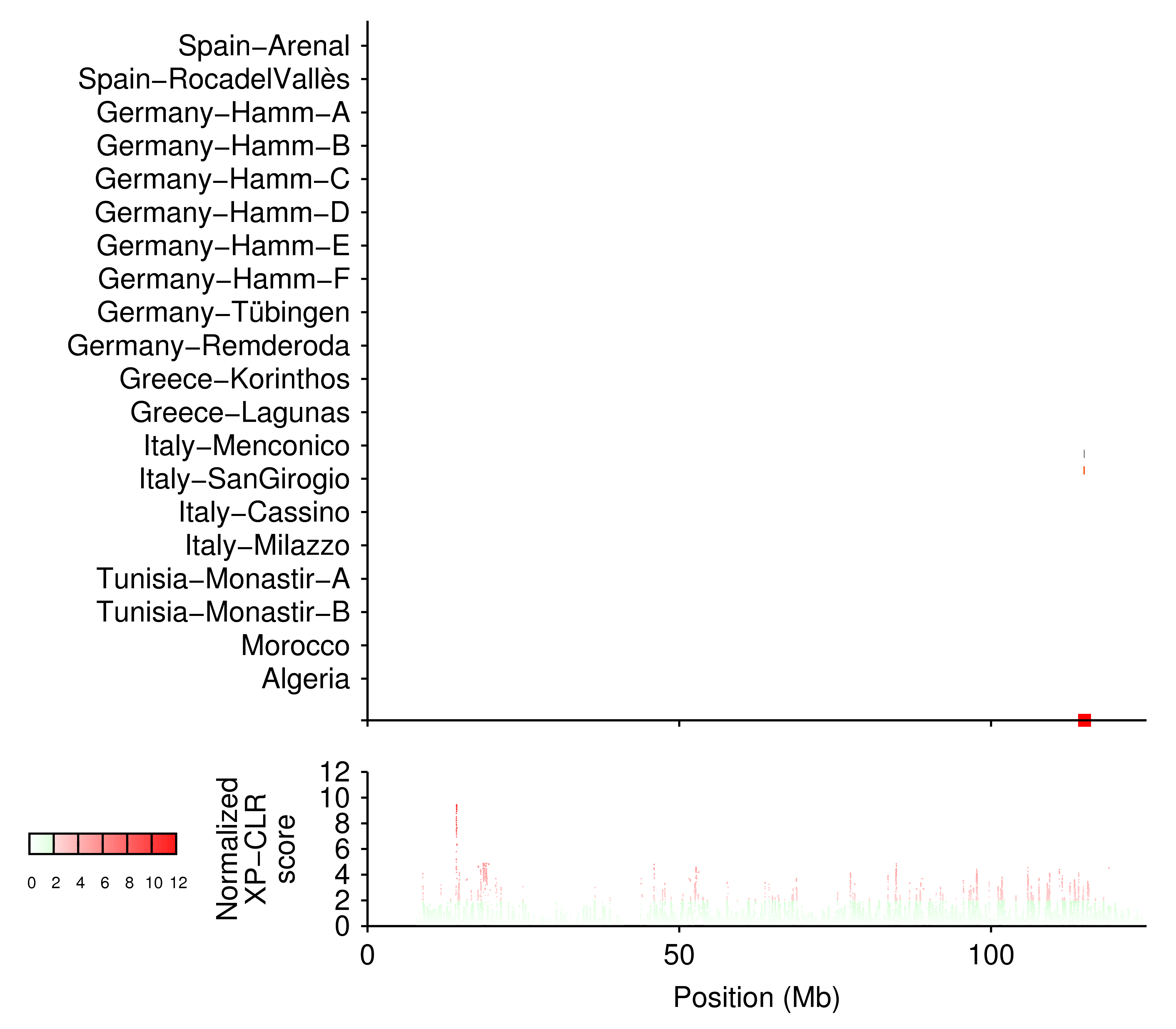}}
\end{center}
\caption{
{\bf Introgression scans of chromosome 14.} 
Figure layout and description are otherwise identical to Figure \ref{hybrid-chr1}.
}
\label{hybrid-chr14}
\end{figure*}

\clearpage

\begin{figure*}[!ht]
\begin{center}
\centerline{\includegraphics[scale=1.7]{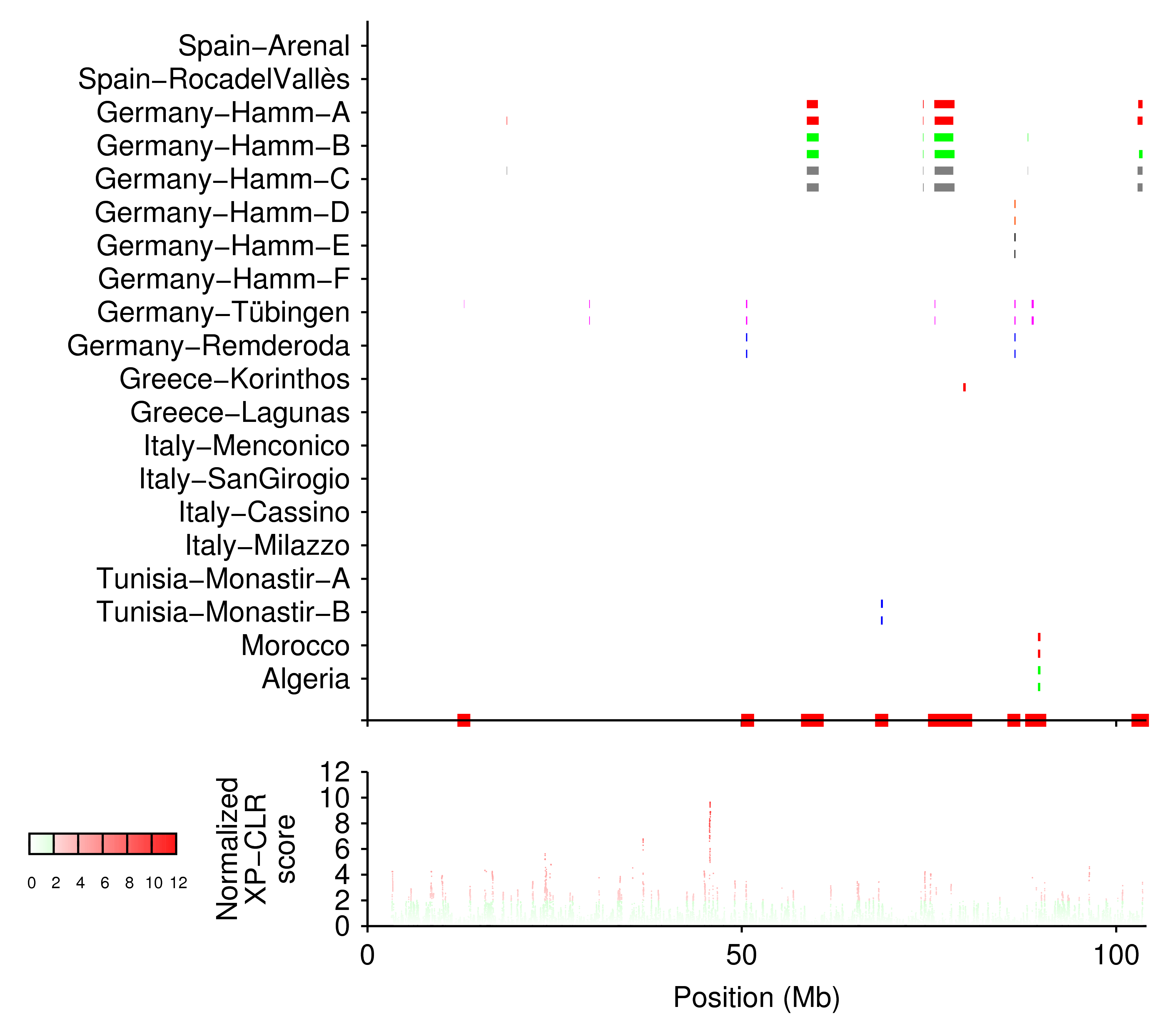}}
\end{center}
\caption{
{\bf Introgression scans of chromosome 15.} 
Figure layout and description are otherwise identical to Figure \ref{hybrid-chr1}.
}
\label{hybrid-chr15}
\end{figure*}

\clearpage

\begin{figure*}[!ht]
\begin{center}
\centerline{\includegraphics[scale=1.7]{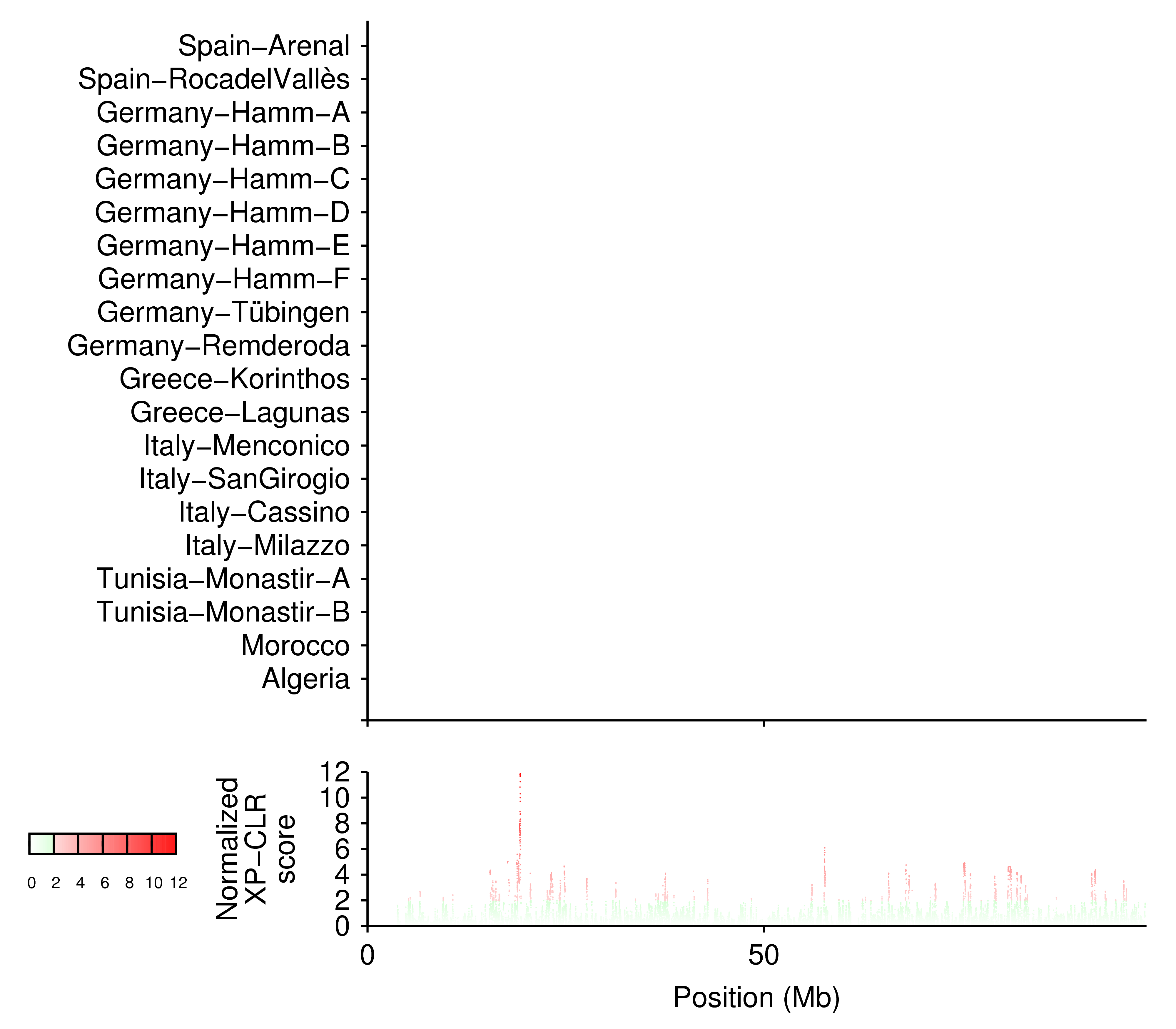}}
\end{center}
\caption{
{\bf Introgression scans of chromosome 16.} 
Figure layout and description are otherwise identical to Figure \ref{hybrid-chr1}.
}
\label{hybrid-chr16}
\end{figure*}

 \clearpage

\begin{figure*}[!ht]
\begin{center}
\centerline{\includegraphics[scale=1.7]{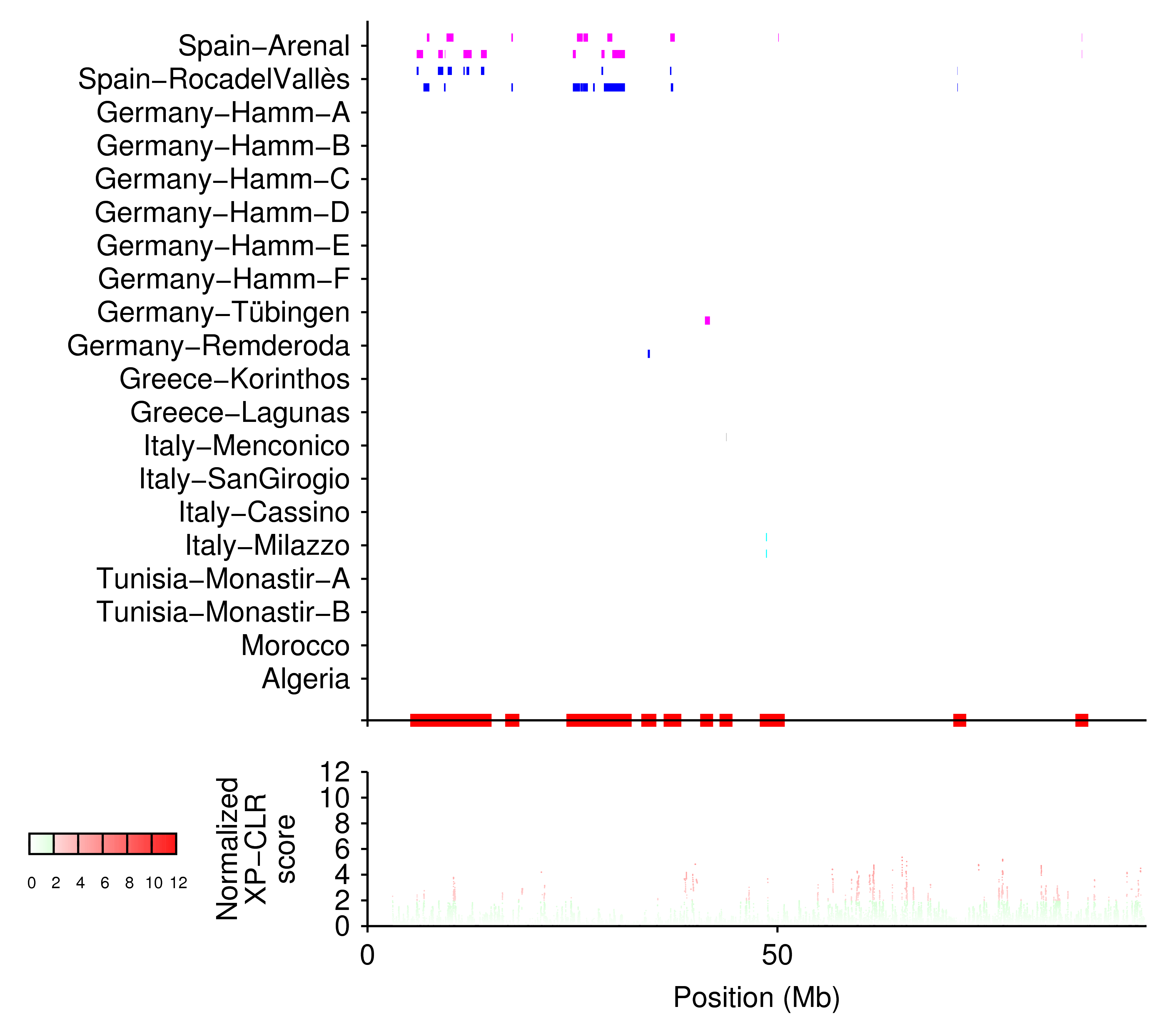}}
\end{center}
\caption{
{\bf Introgression scans of chromosome 17.} 
Figure layout and description are otherwise identical to Figure \ref{hybrid-chr1}.
}
\label{hybrid-chr17}
\end{figure*}

 \clearpage

\begin{figure*}[!ht]
\begin{center}
\centerline{\includegraphics[scale=1.7]{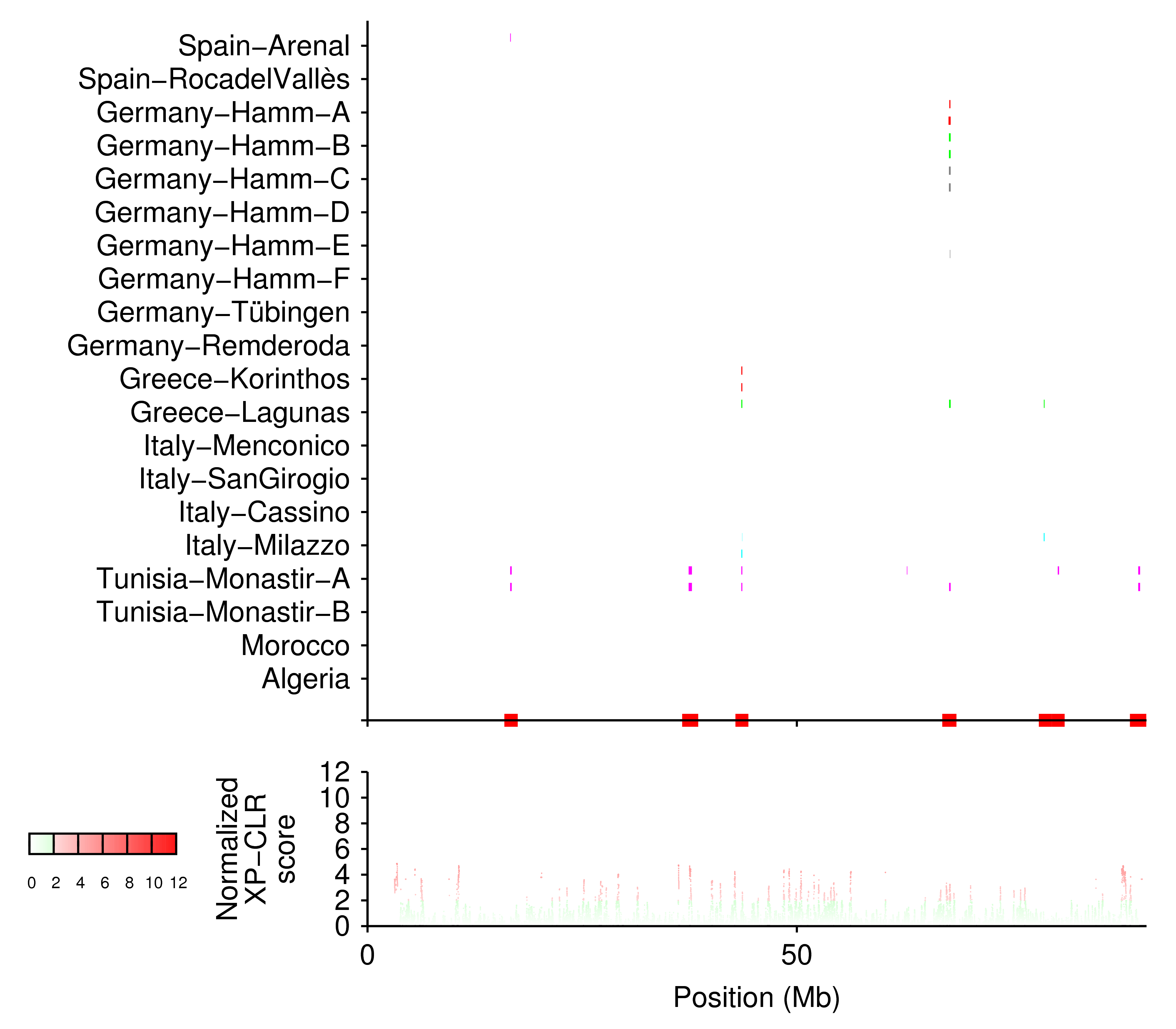}}
\end{center}
\caption{
{\bf Introgression scans of chromosome 18.} 
Figure layout and description are otherwise identical to Figure \ref{hybrid-chr1}.
}
\label{hybrid-chr18}
\end{figure*}

 \clearpage

\begin{figure*}[!ht]
\begin{center}
\centerline{\includegraphics[scale=1.7]{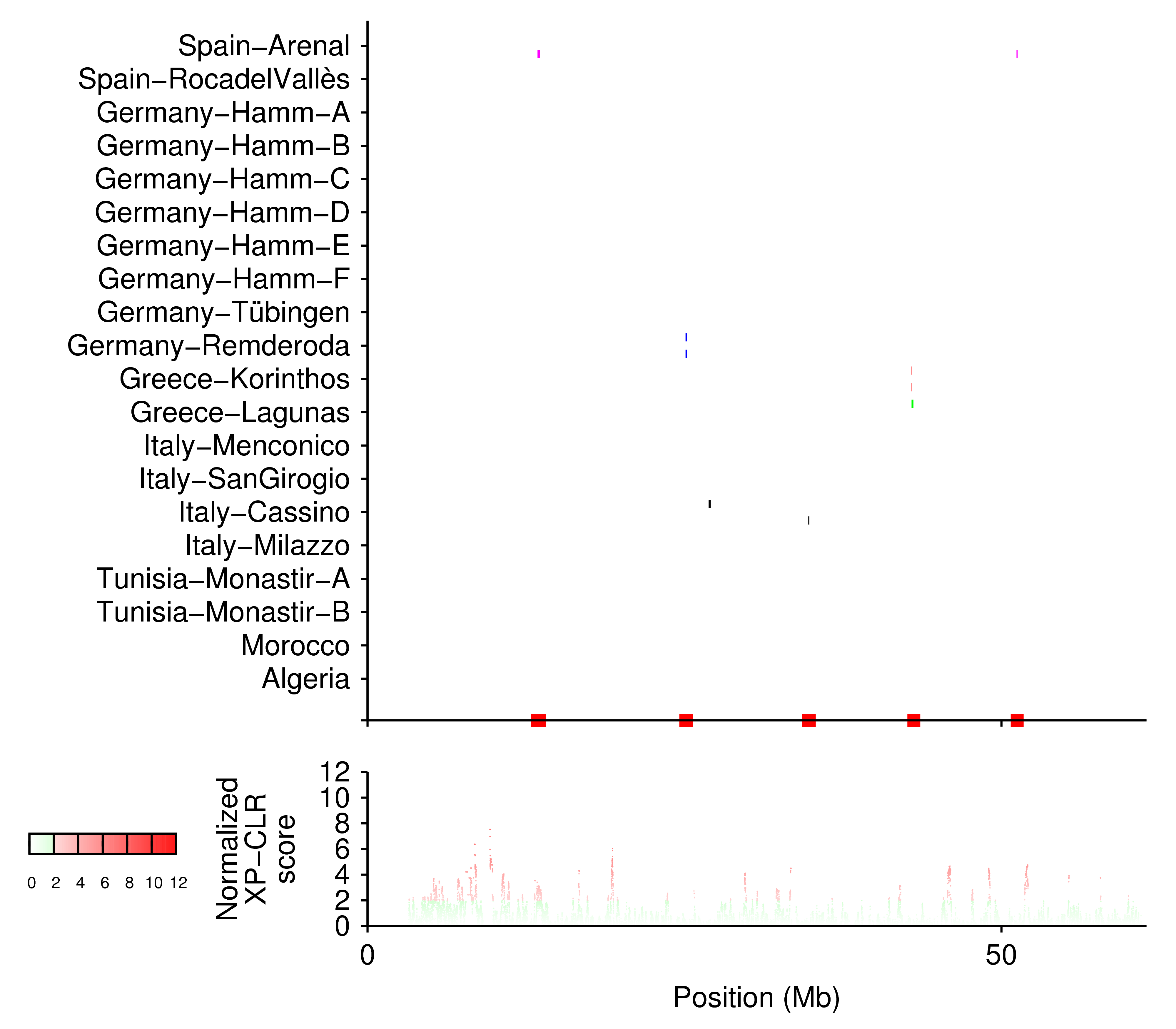}}
\end{center}
\caption{
{\bf Introgression scans of chromosome 19.} 
Figure layout and description are otherwise identical to Figure \ref{hybrid-chr1}.
}
\label{hybrid-chr19}
\end{figure*}

\clearpage

\begin{figure*}[!ht]
\begin{center}
\centerline{\includegraphics[scale=1.7]{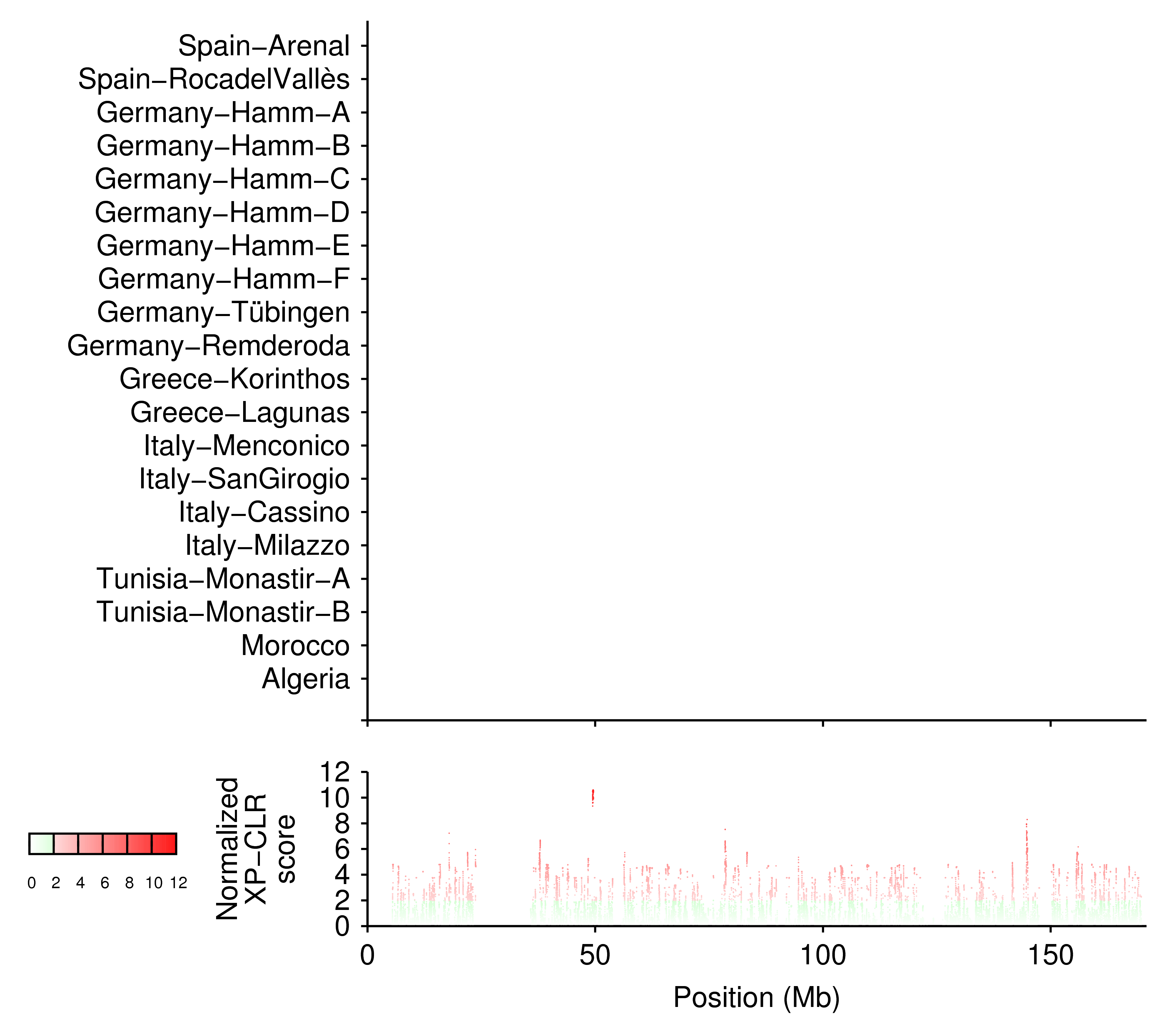}}
\end{center}
\caption{
{\bf Introgression scans of chromosome X.} 
Figure layout and description are otherwise identical to Figure \ref{hybrid-chr1}.
}
\label{hybrid-chrX}
\end{figure*}

\clearpage

\begin{figure*}[!ht]
\begin{center}
\centerline{\includegraphics[scale=0.28]{./charts-si-updated/data-C57BL_6J.UNIQUE.COPY.1-C57BL_6J.UNIQUE.COPY.2-C57BL_6J.UNIQUE.COPY.3-SPRET.EIJ.SANGER.WGS-RN5.REFERENCE.WGS/chr7/KEEP_PARSIMONY_UNINFORMATIVE_ALIGNMENT_SITES:0/FASTPHASE/PHMM2-EXTENDED-2/CHOICE0_KEEP_PARSIMONY_UNINFORMATIVE_CHARACTERS_CHOICES:1/CHOICE1_OPTIMIZE_SUBSTITUTION_MODEL_CHOICES:1/CHOICE2_UNCONSTRAINED_PARENTAL_TREE_OPTIMIZATION_CHOICES:1/BEST/collated-abbreviated.png}}
\end{center}
\caption{
{\bf Results from PhyloNet-HMM scan of chromosome 7 from the reference strain control data set.} 
(a) The posterior decoding probability that a genomic segment had introgressed origin from {\it M. spretus}.
(b) For introgressed genomic segments, the posterior decoding probability of each rooted
gene genealogy is shown for each locus using a heatmap. Each of the 15 possible rooted gene genealogies is shown in a 
different track. A block in a track is shaded
according to the probability of observing the rooted gene genealogy
in the corresponding genomic segment, where
shades are a continuous gradient from white to blue corresponding
to probabilities from $0$ to $1$. 
(c) For non-introgressed genomic segments, the posterior decoding probability of each rooted
gene genealogy is shown for each locus, using
a heatmap similar to panel (b).
(d) The locations of sampled loci are shown.
}
\label{reference-control-chr7}
\end{figure*}

\clearpage

\begin{figure*}[!ht]
\begin{center}
\centerline{\includegraphics[scale=0.28]{./charts-si-updated/data-Yu2097m-Yu2120f-Yu2115m-SPRET.EIJ.SANGER.WGS-RN5.REFERENCE.WGS/chr7/KEEP_PARSIMONY_UNINFORMATIVE_ALIGNMENT_SITES:0/FASTPHASE/PHMM2-EXTENDED-2/CHOICE0_KEEP_PARSIMONY_UNINFORMATIVE_CHARACTERS_CHOICES:1/CHOICE1_OPTIMIZE_SUBSTITUTION_MODEL_CHOICES:1/CHOICE2_UNCONSTRAINED_PARENTAL_TREE_OPTIMIZATION_CHOICES:1/BEST/collated-abbreviated.png}}
\end{center}
\caption{
{\bf Results from PhyloNet-HMM scan of chromosome 7 from the {\it M. m. musculus} control data set.} 
Figure layout and description are otherwise identical to Figure \ref{reference-control-chr7}.
}
\label{mmm-control-chr7}
\end{figure*}

\clearpage

\clearpage

\begin{table*}[!ht]
\centering
\scriptsize
\caption{
{\bf Introgressed genes show significant 
enrichment in olfaction
and other functional categories.
} 
Functional categories are from the
Gene Ontology (GO) \cite{Ashburner2000}.
Statistical testing was performed using
BiNGO \cite{Maere15082005}
(hypergeometric test with Benjamini-Hochberg \cite{BenjaminiHochberg} correction; $\alpha < 0.05$).
(See Materials and Methods for more details.)
}\label{go-terms-for-introgressed-genes}
\vspace{10mm}
\begin{tabular}{rcr}
\hline
GO term ID & Description & Corrected q-value \\
\hline
50911 & detection of chemical stimulus involved in sensory perception of smell & $<$1e-5 \\
4984 & olfactory receptor activity & $<$1e-5 \\
50907 & detection of chemical stimulus involved in sensory perception & $<$1e-5 \\
50906 & detection of stimulus involved in sensory perception & $<$1e-5 \\
9593 & detection of chemical stimulus & $<$1e-5 \\
7608 & sensory perception of smell & $<$1e-5 \\
7606 & sensory perception of chemical stimulus & $<$1e-5 \\
51606 & detection of stimulus & $<$1e-5 \\
7600 & sensory perception & $<$1e-5 \\
4930 & G-protein coupled receptor activity & $<$1e-5 \\
7186 & G-protein coupled receptor signaling pathway & $<$1e-5 \\
60089 & molecular transducer activity & $<$1e-5 \\
4871 & signal transducer activity & $<$1e-5 \\
50877 & neurological system process & $<$1e-5 \\
4888 & transmembrane signaling receptor activity & $<$1e-5 \\
38023 & signaling receptor activity & $<$1e-5 \\
4872 & receptor activity & $<$1e-5 \\
3008 & system process & $<$1e-5 \\
7166 & cell surface receptor signaling pathway & $<$1e-5 \\
7165 & signal transduction & $<$1e-5 \\
44700 & single organism signaling & $<$1e-5 \\
23052 & signaling & $<$1e-5 \\
7154 & cell communication & $<$1e-5 \\
42221 & response to chemical stimulus & $<$1e-5 \\
51716 & cellular response to stimulus & $<$1e-5 \\
71944 & cell periphery & $<$1e-5 \\
5886 & plasma membrane & $<$1e-5 \\
31224 & intrinsic to membrane & $<$1e-5 \\
16021 & integral to membrane & $<$1e-5 \\
50896 & response to stimulus & $<$1e-5 \\
44425 & membrane part & $<$1e-5 \\
50794 & regulation of cellular process & $<$1e-5 \\
44707 & single-multicellular organism process & $<$1e-5 \\
32501 & multicellular organismal process & $<$1e-5 \\
44763 & single-organism cellular process & $<$1e-5 \\
44699 & single-organism process & $<$1e-5 \\
16020 & membrane & $<$1e-5 \\
50789 & regulation of biological process & $<$1e-5 \\
65007 & biological regulation & $<$1e-5 \\
9987 & cellular process & $<$1e-5 \\
5344 & oxygen transporter activity & 5e-4 \\
15671 & oxygen transport & 1e-5 \\
35589 & G-protein coupled purinergic nucleotide receptor signaling pathway & 3e-3 \\
45028 & G-protein coupled purinergic nucleotide receptor activity & 3e-3 \\
1608 & G-protein coupled nucleotide receptor activity & 3e-3 \\
42613 & MHC class II protein complex & 6e-3 \\
15669 & gas transport & .010 \\
31721 & hemoglobin alpha binding & .012 \\
23014 & signal transduction by phosphorylation & .012 \\
50798 & activated T cell proliferation & .016 \\
5833 & hemoglobin complex & .016 \\
44464 & cell part & .030 \\
5623 & cell & .033 \\
4702 & receptor signaling protein serine/threonine kinase activity & .034 \\
19882 & antigen processing and presentation & .037 \\
42611 & MHC protein complex & .040 \\
\hline 
\end{tabular}
\end{table*}


\clearpage

\clearpage








\begin{longtable}{p{0.2\textwidth} p{0.15\textwidth} p{0.12\textwidth} p{0.1\textwidth} p{0.1\textwidth}}
\caption{
{\bf Location of introgressed genes 
detected by PhyloNet-HMM.} 
Results based on a posterior decoding probability
cutoff of 95\%. Coordinates are relative to the
GRCm38.p2 reference genome. 
}\label{introgressed-gene-locations}\\
\hline
\textbf{Ensembl} & \textbf{MGI} &  & \textbf{Start} & \textbf{End} \\
\textbf{ID} & \textbf{Symbol} & \textbf{Chromosome} & \textbf{Coordinate} & \textbf{Coordinate} \\
\hline
\endfirsthead
\multicolumn{5}{c}%
{\tablename\ \thetable\ -- \textit{Continued from previous page}} \\
\hline
\textbf{Ensembl} & \textbf{MGI} &  & \textbf{Start} & \textbf{End} \\
\textbf{ID} & \textbf{Symbol} & \textbf{Chromosome} & \textbf{Coordinate} & \textbf{Coordinate} \\
\hline
\endhead
\hline \multicolumn{5}{r}{\textit{Continued on next page}} \\
\endfoot
\hline
\endlastfoot
ENSMUSG00000051951 & Xkr4 & 1 & 3205901 & 3671498 \\
ENSMUSG00000088390 & Gm26309 & 1 & 3678880 & 3678943 \\
ENSMUSG00000089420 & Gm26010 & 1 & 3783737 & 3783798 \\
ENSMUSG00000026080 & Chst10 & 1 & 38863877 & 38898160 \\
ENSMUSG00000067604 & Nms & 1 & 38939149 & 38950278 \\
ENSMUSG00000026078 & Pdcl3 & 1 & 38987814 & 38997236 \\
ENSMUSG00000026077 & Npas2 & 1 & 39193731 & 39363234 \\
ENSMUSG00000026074 & Map4k4 & 1 & 39900913 & 40026310 \\
\end{longtable}

\clearpage

\begin{longtable}{p{0.2\textwidth} p{0.1\textwidth} p{0.12\textwidth} p{0.1\textwidth} p{0.1\textwidth}|}
\caption{
{\bf Location of introgressed tracts
detected by PhyloNet-HMM.} 
Results based on a posterior decoding probability
cutoff of 95\%. Coordinates are relative to the
GRCm38.p2 reference genome. 
}\label{introgressed-tract-locations}\\
\hline
\textbf{} & \textbf{Haploid} & \textbf{} & \textbf{Start} & \textbf{End} \\
\textbf{Sample} & \textbf{Genome} & \textbf{Chromosome} & \textbf{Coordinate} & \textbf{Coordinate} \\
\hline
\endfirsthead
\multicolumn{5}{c}%
{\tablename\ \thetable\ -- \textit{Continued from previous page}} \\
\hline
\textbf{} & \textbf{Haploid} & \textbf{} & \textbf{Start} & \textbf{End} \\
\textbf{Sample} & \textbf{Genome} & \textbf{Chromosome} & \textbf{Coordinate} & \textbf{Coordinate} \\
\hline
\endhead
\hline \multicolumn{5}{r}{\textit{Continued on next page}} \\
\endfoot
\hline
\endlastfoot
Spain-Arenal & 1 & 4 & 112010036 & 114305637 \\
Spain-Arenal & 1 & 5 & 97598394 & 98039561 \\
Spain-Arenal & 1 & 6 & 71257067 & 71360763 \\
Spain-Arenal & 1 & 7 & 103905178 & 104153889 \\
Spain-Arenal & 1 & 7 & 106687970 & 107293973 \\
Spain-Arenal & 1 & 10 & 22906537 & 23268188 \\
Spain-Arenal & 1 & 10 & 88449176 & 88885679 \\
Spain-Arenal & 1 & 12 & 21178749 & 21535128 \\
\end{longtable}